# Novel Strongly Correlated Europium Superhydrides

Dmitrii V. Semenok,[1] Di Zhou,[2] Alexander G. Kvashnin,[1] Xiaoli Huang,[2,*] Michele Galasso,[1] Ivan A. Kruglov,[3,4] Anna G. Ivanova,[5] Alexander G. Gavrilyuk,[5,6] Wuhao Chen,[2] Nikolay V. Tkachenko,[7] Alexander I. Boldyrev,[7] Ivan Troyan,[5] Artem R. Oganov,[1,*] and Tian Cui[8,2]

[1] Skolkovo Institute of Science and Technology, Skolkovo Innovation Center, 3 Nobel Street, Moscow 143026, Russia
[2] State Key Laboratory of Superhard Materials, College of Physics, Jilin University, Changchun 130012, China
[3] Moscow Institute of Physics and Technology, 9 Institutsky Lane, Dolgoprudny 141700, Russia
[4] Dukhov Research Institute of Automatics (VNIIA), Moscow 127055, Russia
[5] Shubnikov Institute of Crystallography, Federal Scientific Research Center Crystallography and Photonics, Russian Academy of Sciences, 59 Leninskii pr-t, Moscow 119333, Russia
[6] IC RAS Institute for Nuclear Research, Russian Academy of Sciences, Moscow 117312, Russia
[7] Department of Chemistry and Biochemistry, Utah State University, 0300 Old Main Hill, Logan, Utah, 84322-0300, USA
[8] School of Physical Science and Technology, Ningbo University, Ningbo 315211, China

**Corresponding Authors**

Dr. X. Huang, e-mail: huangxiaoli@jlu.edu.cn, Prof. A. Oganov, e-mail: a.oganov@skoltech.ru

## Abstract

We conducted a joint experimental–theoretical investigation of the high-pressure chemistry of europium polyhydrides at pressures of 86–130 GPa. We discovered several novel magnetic Eu superhydrides stabilized by anharmonic effects: cubic $EuH_9$, hexagonal $EuH_9$ and an unexpected cubic ($Pm\bar{3}n$) clathrate phase $Eu_8H_{46}$. Monte Carlo simulations indicate that cubic $EuH_9$ has an antiferromagnetic ordering with $T_N$ up to 24 K, whereas hexagonal $EuH_9$ and $Pm\bar{3}n$-$Eu_8H_{46}$ possess a ferromagnetic ordering with $T_C$ = 137 and 336 K, respectively. The electron–phonon interaction is weak in all studied europium hydrides, and their magnetic ordering excludes $s$-wave superconductivity, except, perhaps, distorted pseudohexagonal $EuH_9$. The equations of state predicted within the DFT+U approach ($U$–$J$ were found within linear response theory) are in close agreement with the experimental data. This work shows the great influence of the atomic radius on symmetry-breaking distortions of the crystal structures of superhydrides and on their thermodynamic stability.

**Table of Content**

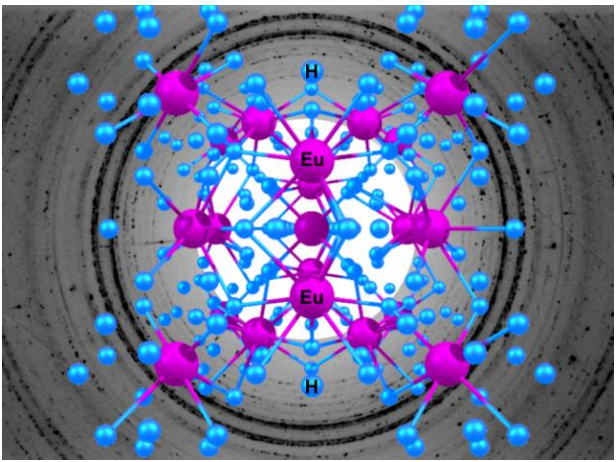

The high-pressure synthesis of new compounds of metal hydrides is a rapidly developing field, in view of unusual chemistry and record high-$T_C$ superconductivity of some hydrides. Some extraordinary compounds, such as $CSH_x$,[1] $LaH_{10}$,[2-4] $YH_6$ and $YH_9$,[5-6] $ThH_9$ and $ThH_{10}$,[7] $UH_7$,[8] $CeH_9$,[9-10] $PrH_9$,[11] and $NdH_9$[12] have already been synthesized. Some of these hydrides, for example, superhydrides of yttrium,[5-6] lanthanum,[3-4] and thorium,[7] demonstrate remarkable superconductivity. However, magnetic ordering emerging in the hydrides of many *f*-metals at low temperatures partly or completely destroys their superconductivity (polyhydrides of praseodymium $PrH_9$,[11] neodymium $NdH_9$[12]). As we have recently shown for $NdH_9$,[12] taking into account the "spin splitting"[13] of the electron bands near the Fermi level rules classical superconductivity out because of the high energy gap that electrons need to overcome for s-pairing. Nevertheless, the possibility of d-pairing remains in such materials.

On the other hand, studies of lanthanoid superhydrides are motivated by the need of further developments in the description of strongly correlated systems. In the studies of the Pr–H[11] and Nd–H[12] systems, we have found that the convex hulls obtained with and without the inclusion of spin–orbit coupling (SOC), as well as with or without magnetism and the Hubbard-like correction ($U$–$J$)[14] differ significantly. In this work, we have continued the study of high-pressure chemistry of lanthanoid hydrides using the Eu-H system.

At normal conditions, europium has the electronic configuration $[Xe]4f^76s^2$, unusually low density and melting point, and *bcc* lattice with relatively large cell parameter $a$ = 4.581 Å.[15] Europium is an anomalous lanthanoid: it is usually divalent, and has an abnormally high atomic radius of ~200 pm, about 10% larger than Y, Sm, and Gd. Under pressure, it undergoes a series of phase transitions ($bcc \rightarrow hcp \rightarrow C2/c \rightarrow Pnma$[16]) accompanied by the emergence of superconductivity above 80 GPa[17] at 1.8–2.75 K due to the conversion of the Eu atoms from the divalent to the trivalent state with zero magnetic moment. Europium readily reacts with hydrogen, forming *Pnma*-$EuH_2$. With an excess of hydrogen under pressure the reaction proceeds further, and trivalent tetragonal $EuH_{3-x}$ may be obtained at about 10 GPa.[18-19] In this work, we continue the study of Eu–H system at pressures up to 130 GPa.

## 1. Experiment

To investigate the formation of new chemical compounds in the Eu–H system at high pressures, we loaded three high-pressure diamond anvil cells (DACs #E1-3) with metallic europium and sublimated ammonia borane $NH_3BH_3$ (AB), used as both a source of hydrogen and a pressure transmitting medium. A tungsten gasket was pressed to 20 μm. Additional parameters of the high-pressure diamond anvil cells are listed in Supporting Information Table S1.

Laser heating of the metal Eu sample in the AB medium at 1600–1800 K for 0.1 s yielded a complex mixture of various europium hydrides observed in DACs #E1 and #E3. The experimental X-ray diffraction (XRD) patterns are shown in Figure **1**. The analysis of the original diffraction images demonstrated the existence of a fine-grained phase with a set of reflections (see inset of Figure **1**b) that can be indexed in space group $Pm\bar{3}n$ with $a$ = 5.858 Å and $V$ = 25.13 Å$^3$/f.u. at 130 GPa, which correspond to composition $EuH_{5+x}$, $0 < x < 1$. Similar diffraction patterns have been previously observed in the U–H system[8] (and attributed to β-$UH_3$) and in the investigation of the Eu–H system by Ma et al.,[20] where the authors attributed it to $Pm\bar{3}n$-$EuH_5$.

Two other sets of reflections may be indexed in space groups $Fm\bar{3}m$ or $F\bar{4}3m$ ($a$ = 4.947 Å, $V$ = 30.27 Å$^3$/f.u. at 130 GPa) and $P6_3/mmc$ ($a$ = 3.591 Å, $c$ = 5.509 Å, $V$ = 30.76 Å$^3$/f.u. at 130 GPa), which may be proposed by analogy with the chemistry of the Pr–H[11] and Nd–H[12] systems (Figure **1a,b**). Both compounds have the cell volume close to that of Eu:H = 1:9 composition. At high pressure (120-130 GPa, DAC #E1) hexagonal polyhydride $EuH_9$ is dominant, whereas the cubic modification was found in small amount, while below 90 GPa situation is opposite (DAC #E3, Figure **1c,d**). Changing the pressure in cells #E1 and # E3 allowed us to get the pressure dependence of the cell volume and compare it with the theoretically calculated equations of state (EoS) for the obtained phases (Figures **2**, **3**). As demonstrated below, this comparison makes it possible to determine the hydrogen content in the discovered compounds and confirm our guesses about $EuH_9$. Thus, we found that $F\bar{4}3m$-$EuH_9$ is present in both diamond anvil cells in the pressure range from 86 to 130 GPa, while the formation of $P6_3/mmc$-EuH$_9$ requires higher pressure (see also Figure S2).

Additional laser heating of DACs #E1 and #E2 at 74 GPa under the same conditions (~1600 K, 0.1 s) with consequent registration of the X-ray diffraction at synchrotron radiation facility SPring-8 (Japan, λ = 0.413 Å) left the diffraction pattern almost unchanged. The analysis showed that the samples still contained a mixture of cubic and hexagonal EuH$_9$ and $Pm\bar{3}n$ phase (Supporting Information Figures S2–S3, S12–S15). The reaction products were unevenly distributed over the volume of the sample: $P6_3/mmc$-EuH$_9$ (Supporting Information Figures S2 and S14, high granularity) or $Pm\bar{3}n$-EuH$_{5+x}$ (Supporting Information Figures S12 and S15, low granularity) may dominate other compounds. In DAC #E2, after compressing to 89 GPa, we mainly found $Pm\bar{3}n$-EuH$_{5+x}$ (Supporting Information Figure S15). Thus, additional laser heating neither improved the purity of the studied mixture nor yielded any new europium hydrides. The formation and stability of $Pm\bar{3}n$-EuH$_{5+x}$ was confirmed at least to 89 GPa.

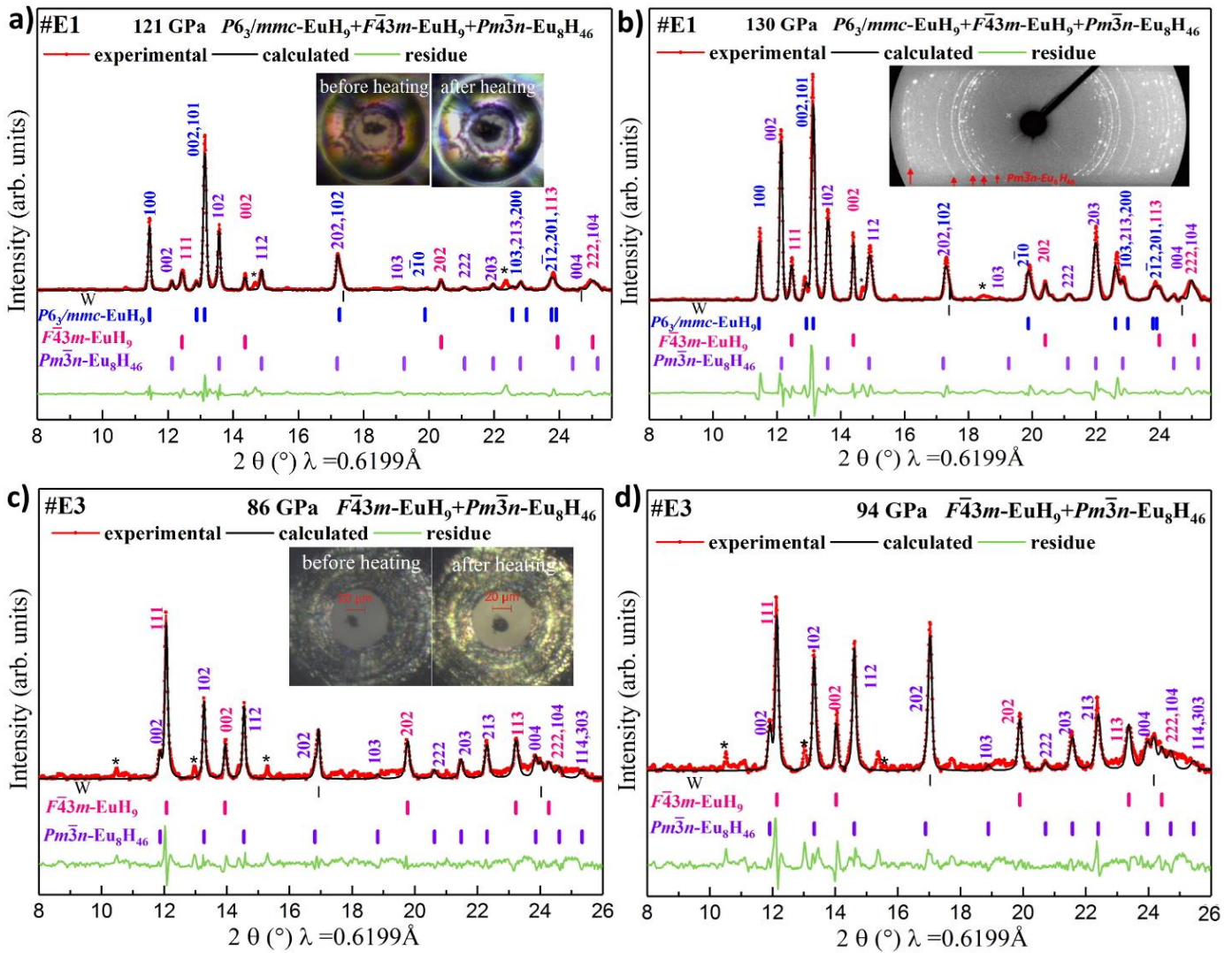

**Figure 1.** Experimental XRD patterns and Le Bail refinements of Eu hydrides. Le Bail refinements of $F\bar{4}3m$-EuH$_9$, $P6_3/mmc$-EuH$_9$, and $Pm\bar{3}n$-Eu$_8$H$_{46}$ at **(a)** 121 GPa and **(b)** 130 GPa. Inset: experimental XRD pattern (cell #E1). The experimental data, model fit for the structure, and residues are shown in red, black, and green, respectively. Unidentified reflections (possibly, $I4/mmm$-EuH$_4$) are indicated by asterisks. The reliable parameters for $F\bar{4}3m$-EuH$_9$, $P6_3/mmc$-EuH$_9$, and $Pm\bar{3}n$-Eu$_8$H$_{46}$ at 121 GPa are Rp = 21.4% and Rwp = 35.6%; at 130 GPa, Rp = 25.3% and Rwp = 34.6%. Le Bail refinements of $F\bar{4}3m$-EuH$_9$ and $Pm\bar{3}n$-Eu$_8$H$_{46}$ at **(c)** 86 GPa and **(d)** 94 GPa of cell #E3. Insets of Figures 1a and 1c show the sample chamber before and after heating of cell #E1 and #E3, respectively.

## 2. Crystal Structure Search

In order to more accurately establish the composition and compute various physical properties of the synthesized phases, we performed a variable-composition search for stable compounds in the Eu–H system at pressures of 50, 100, 130, and 150 GPa using the Universal Structure Predictor: Evolutionary Xtallography (USPEX) algorithm.[21-23] The spin–orbit coupling (SOC) and Hubbard-like correction term ($U$–$J$) were not taken into account in the evolutionary search, but were included when plotting the convex hulls (Figure 1). The $U$–$J$ values were found for all novel europium hydrides using the first-principles linear response approach.[24] It was possible to compute $U$–$J$ using VASP[25-27] by three single-point calculations. The obtained results (Supporting Information Table S4, Figure S4) are close to $U$–$J$ = 4.5–5 eV for all studied compounds.

To construct the convex hulls, we used the experimental XRD data from the previous studies of phase transition of metallic europium. Bi et al.[16] have shown that at 75–92 GPa europium has *Pnma* crystal lattice, with $a$ = 4.977 Å, $b$ = 4.264 Å, $c$ = 2.944 Å, $V$ = 15.62 Å$^3$/atom at 75 GPa. Numerical simulations show that the unit cell volume $V_{Pnma\text{-}Eu}$ shows almost no dependence on the value of $U$–$J$ in the range 0–5 eV and on SOC, but the pseudopotential used in calculations strongly affects this volume: $V_{Eu}$ = 15.82 and 13.49 Å$^3$/atom for 17-electron and 8-electron pseudopotentials, respectively. Because of much better agreement with experiment, Eu pseudopotential with 17 valence electrons was used in all further calculations.

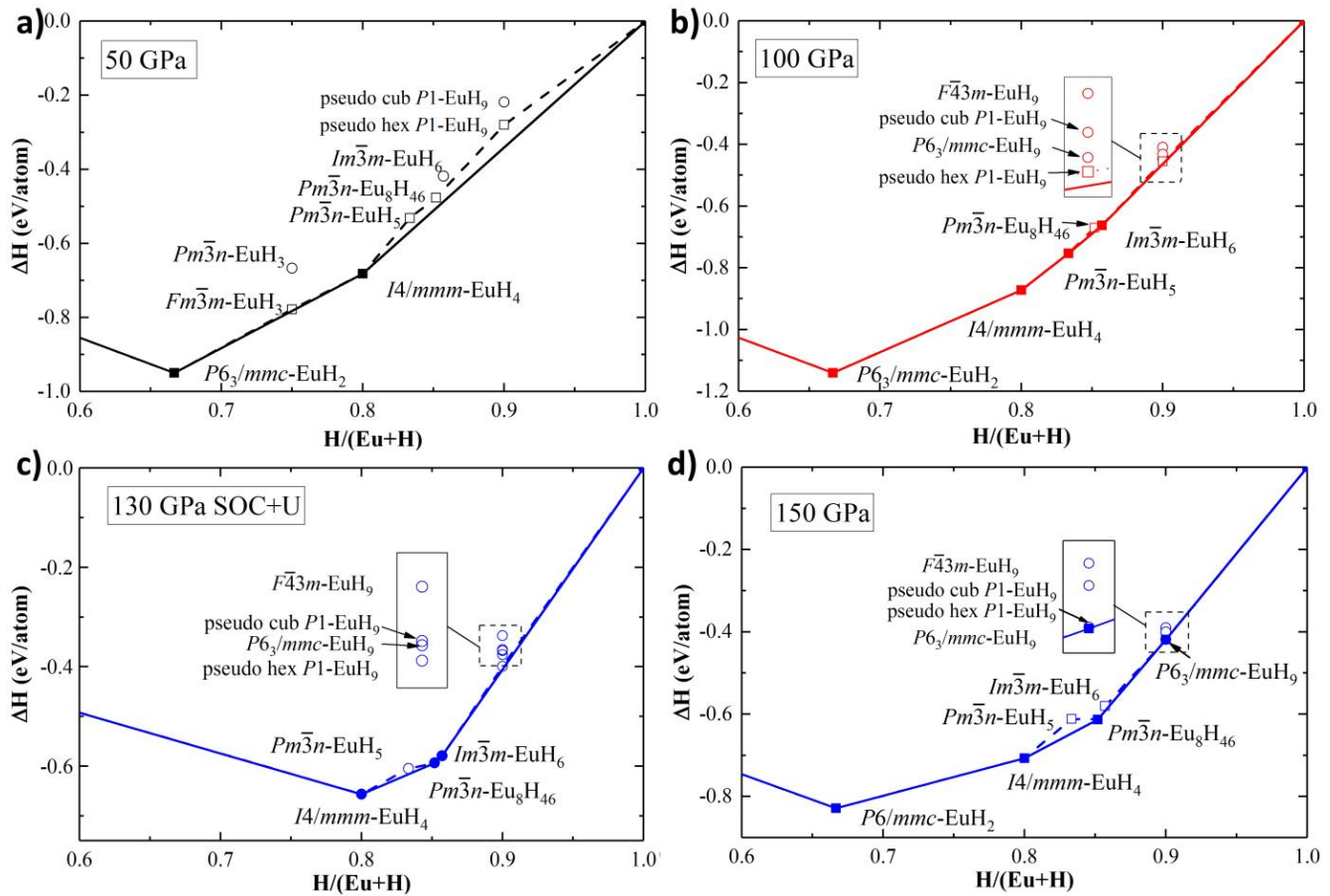

**Figure 2.** Calculated convex hulls of the Eu–H system at **(a)** 50 GPa, **(b)** 100 GPa, and **(d)** 150 GPa without SOC and $U$-$J$ = 0, and **(c)** at 130 GPa with SOC and specific $U$–$J$.

*I4/mmm*-EuH$_4$, a compound from the group of previously found hydrides *I4/mmm*-XH$_4$ (X = Ce, Pr, Nd), and $Im\bar{3}m$-EuH$_6$[20] are placed on the convex hull of the Eu–H system in the experimental range of pressure from 100 to 130 GPa (Figure 2). At low pressures of ~50 GPa, only *I4/mmm*-EuH$_4$ is expected, whereas higher polyhydrides are unstable. The inclusion of SOC and $U$–$J$, as well as the analysis of the entropy factor (Supporting Information, Table S9, Figure S10), do not change the set of stable Eu–H phases, although at 130 GPa and 0 K EuH$_9$ is slightly (20 meV/atom) above the convex hull. However, the analysis of the equation of state (Figure 3) points to several deviations of $V(P)$ of the predicted europium hydrides from the experimental volumes.

First, the predicted unit cell volume of $Pm\bar{3}n$-EuH$_5$ is 22.8–23.9 Å$^3$/f.u. at 130 GPa (without SOC, $U-J = 0$ eV and with SOC and $U-J = 5$ eV, respectively), whereas the experimental value is 25.13 Å$^3$/f.u., about 1-2 Å$^3$ higher. The situation holds for other $U-J$=0–7 eV. To solve the problem, get closer to the experimental volume and find a better structural solution, we increased the hydrogen content $x$ in $Pm\bar{3}n$-EuH$_{5+x}$, performing two USPEX searches at 130 GPa with fixed ratios of Eu:H = 1:6 and 8:46 (1:5.75, Supporting Information Table S10). The latter composition was proposed by analogy with the well-known $Pm\bar{3}n$ phases of Ge, Si and Sn clathrates of alkali and alkaline earth metals, such as K$_4$Ge$_{23}$, Rb$_4$Ge$_{23}$, Ba$_8$Si$_{46}$, and so forth (so-called Zintl clathrates[28]). For the former composition, the best solution corresponds to the recently synthesized $Im\bar{3}m$-EuH$_6$[20]; however, we did not detect this compound in our experiments.

As a result of the second USPEX search, $Pm\bar{3}n$-Eu$_8$H$_{46}$ was found to be the most stable phase with a cell volume of 25.21 Å$^3$/Eu. This structure features H$_{24}$ cages around each Eu atom (among 24 hydrogens, 8 are closest, with Eu-H distances below 1.87 Å). Among the 46 hydrogen atoms in the unit cell, 16 form pairs that can be described as stretched H$_2$ molecules ($d$(H-H) = 0.99 Å vs 0.74 in the free H$_2$ molecule) and the remaining 30 hydrogen atoms can be described as single atoms (their shortest H-H distance is 1.33 Å at 130 GPa). This phase corresponds to the experimental XRD and EoS, lies close to the convex hull with and without SOC and $U-J$ (Figure 2), and can explain the experimental data for "$Pm\bar{3}n$-EuH$_5$" phase (Figure 3) obtained by Ma et al.[20]

Second, the ideal $P6_3/mmc$-EuH$_9$ is out of the thermodynamic stability area (e.g. Figure 2c). It has been shown recently that a similar behavior is observed in the Y–H system[6], where ideal $P6_3/mmc$-YH$_9$ is also unstable thermodynamically and transforms to stable $P1$-Y$_4$H$_{36}$ which has almost the same XRD pattern (see Figure S1). Using this structure as a prototype, we calculated the enthalpy of similar $P1$-Eu$_4$H$_{36}$ (which may be relaxed and symmetrized to a pseudohexagonal $Cmcm$-EuH$_9$) and found that this distorted structure (Supporting Information Figure S1) is more stable than the ideal $P6_3/mmc$-EuH$_9$. Similarly, $F\bar{4}3m$-EuH$_9$ distorts to form more stable pseudocubic $P1$-EuH$_9$.

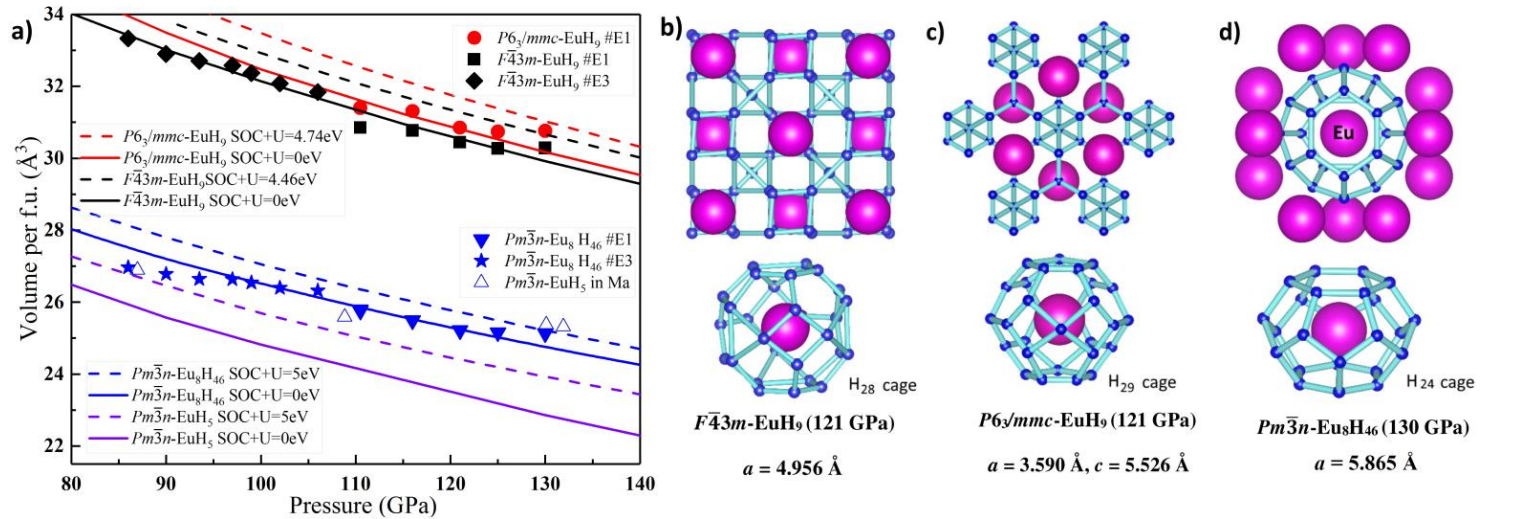

**Figure 3.** Equations of state, crystal structures, and magnetization of the synthesized Eu–H phases. **(a)** Pressure dependence of volumes of europium hydrides. The theoretical results taking into account SOC are shown by solid ($U-J = 0$ eV) and dash-dot ($U-J = 5$ eV) lines. The experimental data for EuH$_5$ from Reference[20] are shown by hollow squares. Crystal structures of ideal $F\bar{4}3m$-EuH$_9$ **(b)**, and $P6_3/mmc$-EuH$_9$ **(c)** at 121 GPa, and $Pm\bar{3}n$-Eu$_8$H$_{46}$ **(d)** at 130 GPa. Purple and blue spheres represent Eu and H atoms, respectively.

The calculated unit cell volume of $Fm\bar{3}m$-EuH$_{10}$ is 31.8–32.4 Å$^3$/f.u. with $U-J = 0$ and 5 eV, respectively (Supporting Information Figure S7), at 130 GPa, which is 1–2 Å$^3$ higher than what we found in experiment for cubic

EuH$_x$ phase (Figures **1, 3a**). Therefore, the existence of $Fm\bar{3}m$-EuH$_{10}$, isostructural to previously found $Fm\bar{3}m$-LaH$_{10}$, is not confirmed by the current experimental data.

It is curious to track how a series of stable europium hydrides changes with pressure. At 50 GPa, all europium polyhydrides except EuH$_4$ are metastable (Figure **2a**). An increase in pressure to 100 GPa stabilizes a wide range of polyhydrides, including the above described $Pm\bar{3}n$-Eu$_8$H$_{46}$, $Im\bar{3}m$-EuH$_6$, as well as pseudocubic $P1$-EuH$_9$ and pseudohexagonal $P1$-EuH$_9$ which is close to $Cmcm$ (Figure **2b**). Further increasing pressure to 150 GPa leads to the stabilization of higher-symmetry structures, and distortions of ideal $F\bar{4}3m$-EuH$_9$ and $P6_3/mmc$-EuH$_9$ cease to play a significant role (Figure **2d**). An increase of temperature to 2000 K leads to the destabilization of EuH$_9$ in favor of EuH$_6$, which may explain the unsuccessful experimental attempts to change the phase composition by additional laser heating (Supporting Information Figure S10). Anharmonic calculations based on molecular dynamics and machine learning potentials of interatomic interactions (see Supporting Information for details) show that ideal $P6_3/mmc$-EuH$_9$, $F\bar{4}3m$-EuH$_9$ and $Pm\bar{3}n$-Eu$_8$H$_{46}$ are dynamically stable, although in the harmonic approximation they should undergo distortions (Supporting Information Figures S26, S29-31). Having predicted the stable compositions and structures of europium hydrides, we now want to get insight into their nature.

## 3. Superatomic Orbitals and Chemical Bonding

The description of electronic structure in terms of delocalized canonical orbitals is almost completely devoid of an explicit chemical bonding picture of a given system. To understand chemical bonding in europium polyhydrides, we first analyzed model clusters (blocks separated from the infinite crystal lattice) via the adaptive natural density partitioning (AdNDP) algorithm.[29] The AdNDP is a localization technique that follows general ideas of the natural bond orbital (NBO) analysis proposed by Weinhold and Landis.[30] The main advantage of this method is a possibility to represent a chemical bonding pattern in terms of both Lewis bonding elements (lone pairs, two-center two-electron (2c–2e) bonds) and delocalized bonding elements ($n$c–2e bonds) by partial diagonalization of the one-body reduced density matrix. This technique is widely used in materials science and in describing various Zintl clusters.[31-33] The chemical bonding analysis can offer insights on the reasons of stability, electronic properties, magnetism, and chemical activity of investigated systems. It has been shown that the chemical bonding analysis of cluster models qualitatively agrees with the analysis of solid state structures.[34-37] In this work, we used EuH$_{27}^{18+}$ as a model of $P6_3/mmc$-EuH$_9$, EuH$_{24}^{17+}$ as a model of $Pm\bar{3}n$-Eu$_8$H$_{46}$, and EuH$_{28}^{19+}$ as a model of $F\bar{4}3m$-EuH$_9$ (Figure **4**). Charges were chosen so as to adjust the overall number of electrons per Eu atom in a studied fragment and consider those electrons shared between the cluster and the neighboring atoms.

Starting our search from localized bonding elements such as 2c–2e, 3c–3e, and so forth, we found no highly occupied localized bonds in the present structures. Due to the absence of highly occupied localized bonding elements, the model clusters could be completely described in terms of fully delocalized canonical molecular orbitals (MOs). As was expected, for such sphere-like clusters, the canonical MOs mimic the spherical harmonics forming so-called superatomic orbitals.[38] Further analysis of the superatomic orbitals of corresponding clusters reveals that unpaired electrons are located mostly on the F-type orbitals (Figure **4**a–f), whereas the S-, P-, and D-type orbitals are doubly occupied.

Following the analysis of the model clusters, we performed the solid-state AdNDP (SSAdNDP) calculations for ideal cubic and hexagonal EuH$_9$ and Zintl-like clathrate $Pm\bar{3}n$-Eu$_8$H$_{46}$. Such analysis was first implemented for polyhydrides in the current work. The SSAdNDP revealed the same bonding picture as was found in the model clusters. A cognate pattern could also be seen in the comparison of the electron localization function (ELF) plots of the solid-state structures and model clusters (Supporting Information Figure S18). The corresponding occupancies of valence superatomic orbitals are presented in Supporting Information Table S6. For all europium polyhydride structures, there are seven almost ideally (1.00 |$e$|, where $e$ is the elementary charge) singly occupied superatomic F-orbitals, which are responsible for the magnetic properties. The corresponding $n$c–2e occupancy numbers (where $n$ is the number of hydrogen atoms in the cage) of superatomic D- and P-orbitals are 1.72–1.39 |$e$| for $P6_3/mmc$-EuH$_9$,

1.17–1.09 |e| for $Pm\bar{3}n$-Eu$_8$H$_{46}$, and 1.57–1.38 |e| for $F\bar{4}3m$-EuH$_9$. Due to delocalization of conducting electrons, which is common for metals,[39] these values are about 0.3–0.9 |e| lower than those we found for clusters.

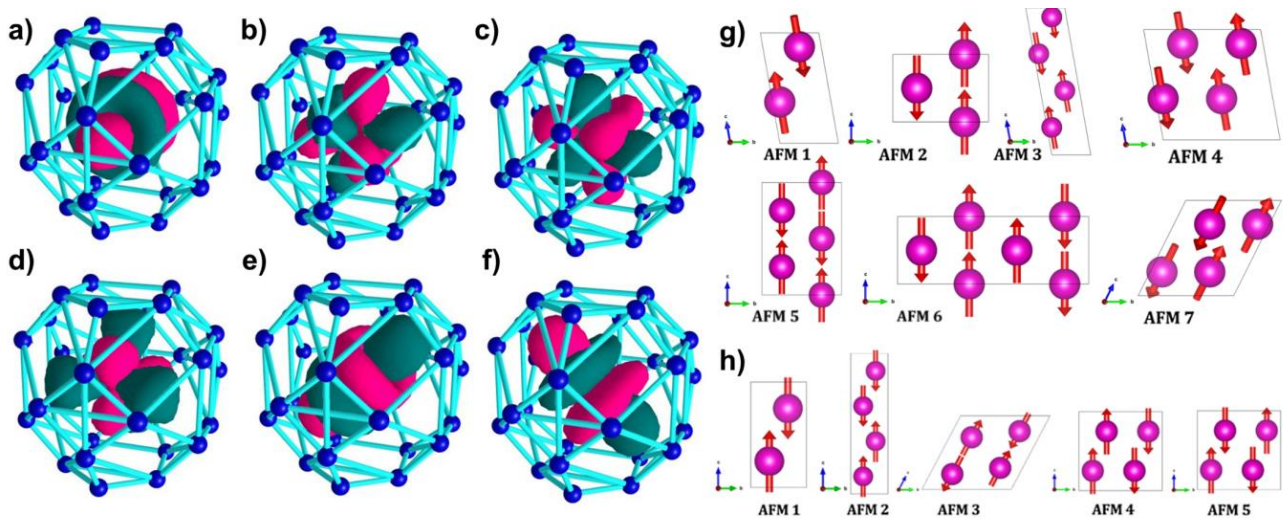

**Figure 4.** (a–f) Singly occupied F-type superatomic orbitals of EuH$_{28}$$^{19+}$ (cluster model of $F\bar{4}3m$-EuH$_9$). (g) Seven trial antiferromagnetic configurations for $F\bar{4}3m$-EuH$_9$ (ferromagnetic is trivial) and (h) five trial antiferromagnetic configurations for $P6_3/mmc$-EuH$_9$ used in the calculations.

To understand the charge distribution in the studied structures, Bader charge analysis was performed.[40] In all structures, the Eu atom bears a positive charge of +1.15 to +0.89 |e|, whereas a partially negative charge of –0.31 to –0.08 |e| was found for all hydrogen atoms. This behavior is common for metal hydrides, similar results have been found in previous works. The average Bader charges are presented in Supporting Information Table S18.

Both cubic and hexagonal EuH$_9$, as well as Eu$_8$H$_{46}$, demonstrate a large magnetic moment (absolute value) in the unit cell (Figure S7), slightly decreasing as pressure rises. At 100 GPa, the magnetic moment of the EuH$_9$ unit cell is 5.96 $\mu_B$ per Eu atom for $P6_3/mmc$ modification, 6.32 $\mu_B$ per Eu atom for $Pm\bar{3}n$-Eu$_8$H$_{46}$ structure, and about 5.88 $\mu_B$ per Eu atom for cubic EuH$_9$. The magnetization arises mostly from the f electrons of the Eu atoms. We observed the same behavior for the corresponding clusters (Supporting Information Table S5): the ground spin state of EuH$_{27}$$^{18+}$ and EuH$_{24}$$^{17+}$ is nonet ($S_{total}$ = 4), whereas for EuH$_{28}$$^{19+}$ the ground spin state is septet ($S_{total}$ = 3).

We performed a comparative analysis of superatomic orbitals of europium hydrides, recently synthesized cubic and hexagonal PrH$_9$,[11] and hexagonal NdH$_9$[12] (Supporting Information Table S11). For PrH$_9$, the shapes of five obtained $n$c–2e bonds do not completely correspond to D- type: two orbitals have the D-type nature, three other $n$c–2e bonds have the F-type nature. This probably happens because of an ambiguous behavior of f and d orbitals of lanthanoids. A similar behavior — mixing of F and D orbitals — was also observed for the molecular clusters of these hydrides and for NdH$_{24}$$^{16+}$ cluster, which has the D-type and F-type orbitals with very close energy.

The analysis of MOs revealed that all clusters could be described within the superatomic concept, according to which the most stable clusters emerge from completely filled superatomic shells. However, CeH$_{29}$$^{20+}$ (Supporting Information Figure S17) has an additional 2P electron that contradicts this concept. We believe that this P-electron plays an important role in the metallic and superconducting ($T_C$ ~ 110 K (unpublished)) properties of cerium polyhydrides as it was previously shown for LaH$_{10}$[41]]. The overall approach of localization of superatomic orbitals in hydrogen cages could also be extended to other superconducting polyhydrides. For example, strong localization of the multicenter bond was found for H$_{32}$ cages present in LaH$_{10}$ and YH$_{10}$ superconductors. As in the case of Ce polyhydride, 2P electrons in these structures were found to have relatively high occupancy numbers. The 2S superatomic orbital, which is present in clusters, is absent in solid state superatoms. We think that these two electrons, localized on the H cage of clusters, may participate in metallic interactions and cannot be localized via the SSAdNDP technique. The detailed comparison of occupancies of the superatomic orbitals in different polyhydrides is presented in Supporting Information Table S17.

## 4. Magnetic Structure

To analyze the magnetic structure of europium hydrides, we studied a series of ferromagnetic (FM) and antiferromagnetic (AFM) configurations. The enthalpies of the optimized structures are presented in Supporting Information Table S11. During the calculation, the ideal cubic and hexagonal structures of hydrides are distorted into more stable low-symmetry pseudocubic and pseudohexagonal modifications. Most of these distorted structures have XRD spectra close to the ideal ones. To simplify magnetic calculations, in several cases we used high symmetry EuH$_9$ prototypes ($F\bar{4}3m$ and $P6_3/mmc$) instead of thermodynamically more stable distorted $P1$-Eu$_4$H$_{36}$. The magnetic structure of the pseudohexagonal $Cmcm$-EuH$_9$ has nevertheless been studied in detail (see Supporting Information).

The most stable collinear magnetic state of $F\bar{4}3m$-EuH$_9$ is AFM 1, which distorts to $P1$ during relaxation. The second most stable configuration is AFM 3, which distorts to $Imm2$. Finally, among the magnetic states that preserve the original ideal $F\bar{4}3m$ symmetry, the most stable is AFM 2. The same is observed for $P6_3/mmc$-EuH$_9$. The most stable collinear magnetic state is AFM 3, but during relaxation the hexagonal structure is distorted and transformed into pseudohexagonal $Cmcm$, which lies on the convex hull (Figure **4**). The second most stable configuration is AFM 4 which distorts to $C2$. Among the magnetic states that preserve the ideal $P6_3/mmc$ symmetry, the most stable is FM. The most stable collinear configuration for Eu$_8$H$_{46}$ is FM. Unlike the EuH$_9$ phases, the magnetic configurations of Eu$_8$H$_{46}$ do not show significant distortions from the ideal $Pm\bar{3}n$ geometry after relaxation.

For each of our three phases, we used the relaxed geometries of the most stable magnetic states to study magnetic anisotropy, by running calculations with SOC. We computed the single-point enthalpy of these structures both in the FM and in one AFM state, with the magnetic moments aligned along different directions ($x$, $y$, $z$, $xy$, $xz$, $yz$, $xyz$). Results are summarized in Supporting Information Table S12. We expect a little magnetic anisotropy for all phases with EuH9 stoichiometry, for which we identified the most stable orientation of the magnetic moments. We do not expect any magnetic anisotropy in $Pm\bar{3}n$-Eu$_8$H$_{46}$ because different orientations of the magnetic moments have almost the same enthalpy up to numerical errors.

To study the Néel and Curie temperatures of $F\bar{4}3m$-EuH$_9$, $P6_3/mmc$-EuH$_9$ and $Pm\bar{3}n$-Eu$_8$H$_{46}$, we modeled the magnetic interaction within the Ising Hamiltonian and obtained the critical temperatures for both EuH$_9$ and Eu$_8$H$_{46}$ in our Heisenberg model from a Monte Carlo simulation as implemented in the VAMPIRE code[42] (Supporting Information Table S14). We obtained the critical temperature $T_N = 24$ K for cubic EuH$_9$, while hexagonal EuH$_9$ and $Pm\bar{3}n$-Eu$_8$H$_{46}$ possess the ferromagnetic ordering with $T_C = 137$ and 336 K, respectively.

## 5. Electron–Phonon Interaction

Strong electron–phonon coupling (EPC) is expected for lanthanides and their polyhydrides under pressure. It has been shown for superhydrides of Ce,[9-10] Pr,[11,43] and Nd[12,44] that the superconducting (SC) properties and the strength of the electron–phonon interactions decrease as the number of f electrons increases and manifestations of magnetism become more pronounced.

The numerical analysis using the UppSC code[12] performed for hexagonal NdH$_9$ (120 GPa)[12] showed that if the effective spin splitting in the band structure $h(k) = [\xi_\uparrow(k) - \xi_\downarrow(k)]/2$, where $\xi(k)$ is the spin-resolved electron dispersion, is about 0.5 eV or higher, superconductivity will be completely suppressed at all $\mu^* \geq 0$. We found that similar spin splitting gap reaches 4–4.5 eV in both EuH$_9$ and Eu$_8$H$_{46}$ (Figure **5a, c, e**), which is 2 orders of magnitude larger than all known SC gaps. This fact excludes singlet superconductivity and $s$-pairing in europium hydrides.

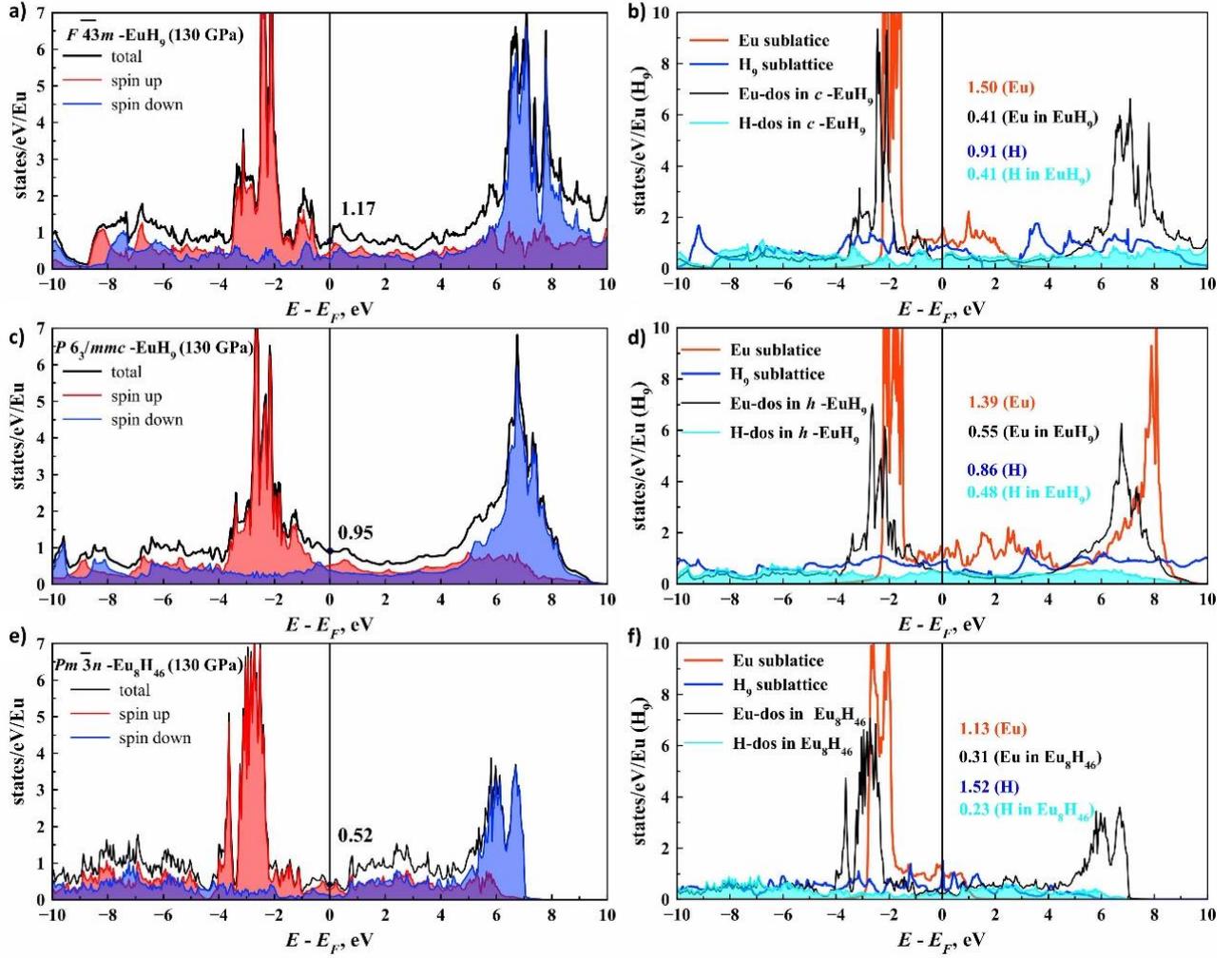

**Figure 5.** Electron density of states (DOS) of europium hydrides at 130 GPa. (a, c, e) Contributions of different spin orientations to the total DOS. (b, d, f) Contributions of the Eu and H atoms to the total DOS. The DOS of the Eu sublattice separated from hydrogen at 130 GPa is shown in red; the DOS of the H$_9$ sublattice separated from the Eu atoms at 130 GPa is shown in blue.

Despite this pessimistic conclusion, we estimated the parameters of the electron–phonon interaction in europium hydrides as if magnetism were absent. The results show that the distorted H sublattice in both modifications of EuH$_9$ could be a superconductor with $T_C$ up to 21 K ($\mu^* = 0.1$), the EPC coefficient $\lambda = 0.61$, and $\omega_{log} = 836$ K for hexagonal EuH$_9$, and $T_C = 27$ K, $\lambda = 0.51$ for cubic EuH$_9$. Compared with LaH$_{10}$ ($T_C \geq 250$ K[3-4]) this indicates an enormous role played by the metal atom (even in the absence of magnetism) and surprisingly different electron-phonon coupling properties of the otherwise very similar lanthanoids. The crucial role of the hydride-forming element and its electronic structure on the $T_C$ of the hydride has been discussed before.[45]

The projection of the total density of states of both EuH$_9$ on the H atoms $N_H(E_F)$ is 0.41–0.48 states/eV/Eu (Figure **5b, d, f**) at 130 GPa. An estimate of the logarithmic frequency made on the basis of the experimental equation of state yields $\omega_{log} = 975–1060$ K for EuH$_9$ at 100 GPa, agreeing with calculations. Both of these factors would be favorable for superconductivity if electron-phonon coupling were stronger and if there were no magnetic ordering. An exception may take place for pseudohexagonal *Cmcm*-EuH$_9$ which does not exhibit magnetism at finite temperatures (see Supporting Information).

Compared to PrH$_9$ and NdH$_9$, f electrons in europium hydrides are localized much deeper (–2 eV) below the Fermi level and $N(E_F)$ is ~0.95–1.17 states/eV/Eu in EuH$_9$, similar to the values for LaH$_{10}$ and YH$_6$, with a high contribution of the H sublattice (~50% of $N(E_F)$). If we consider the metal and hydrogen sublattices in EuH$_9$ separately, disregarding the interaction between them (Figure **5b, d, f**), we will see that the influence of this interaction is completely negative for both $N_H(E_F)$ and $N_{Eu}(E_F)$, which decrease 1.8–2 times. The calculation of the superconducting properties of the formally separated Eu- and H-sublattices of $F\bar{4}3m$-EuH$_9$ at 130 GPa show that the critical temperature of hydrogen ("H$_9$") in the cubic sublattice reaches 150-200 K, while interaction with Eu leads to about 5 times lower $T_C$(EuH$_9$) ~27 K. Thus, the metal lattice plays the role of a stabilizer of metallic hydrogen, but reduces the density of states $N_H(E_F)$ at the Fermi level and the critical temperature of superconductivity.

In conclusion, continuing the studies of the stabilization of metallic hydrogen in compressed metal-hydrogen systems, we synthesized three novel europium superhydrides, with probably slightly distorted structures due to the abnormally large radius of the Eu atom: $P6_3/mmc$-EuH$_9$, $F\bar{4}3m$-EuH$_9$, and an unexpected clathrate phase $Pm\bar{3}n$-Eu$_8$H$_{46}$. The discovered cubic and hexagonal modifications of EuH$_9$ are the new members of the nonahydride family, which currently includes the polyhydrides of Th, U, Y, Ce, Pr, and Nd. Similar to the Nd–H system, all europium hydrides are magnetic and strongly correlated systems with a significant contribution of the SOC interaction. They possess an antiferromagnetic (cubic EuH$_9$) or ferromagnetic (hexagonal EuH$_9$, Eu$_8$H$_{46}$) ordering with $T_{N,C}$ ~ 24–336 K. Although the projection of the DOS on the H sublattice in both phases of EuH$_9$ is similar to that of high-$T_C$ superconducting hydrides, such as LaH$_{10}$ and YH$_6$, electron-phonon coupling is weak. Magnetic order and large effective spin splitting (>4 eV) in the electron band structure of EuH$_9$ makes classical s-wave superconductivity impossible. An exception may take place for pseudohexagonal $Cmcm$-EuH$_9$ which does not exhibit magnetism at finite temperatures.

The chemical bonding analysis was implemented for polyhydrides first time. The analysis points to the contribution of 2P superatomic electrons as an important distinguishing feature in determining high-temperature superconductivity. It was found that the superatomic description of polyhydrides could be extended to other important superconductors, such as LaH$_{10}$ and YH$_{10}$.

Analyzing the whole series of La–Ce–Pr–Nd–Eu superhydrides, we conclude that the role of magnetism is gradually increasing in this row, which leads to suppression of s-wave superconductivity. Most of the higher superhydrides of these metals have an antiferromagnetic ordering, whereas the lower hydrides have a ferromagnetic ordering. Europium hydrides, in addition to Y$_4$H$_{36}$ and BaH$_{12}$, give us yet another indication of the great role of distortions of the ideal hexagonal and cubic structures in their stabilization. Without considering these distortions, theoretical explanation of the thermodynamic stability of europium superhydrides would be impossible.

## Acknowledgments

The authors express their gratitude to the staff of the BL10XU (High Pressure Research) station of SPring-8 synchrotron research facilities, especially to Saori Kawaguchi (JASRI) for the tremendous assistance in the use of the station's equipment before and after the experiment. We also thank the staffs of BL15U station of Shanghai and 4W2 station of Beijing Synchrotron Radiation Facilities.

This work was supported by the National Key R&D Program of China (grant no. 2018YFA0305900), National Natural Science Foundation of China (grant No. 11974133 and 51720105007), National Key Research and Development Program of China (Grant No. 3382016YFB0201204), Program for Changjiang Scholars and Innovative Research Team in University (grant no. IRT_15R23), and National Fund for Fostering Talents of Basic Science (grant no. J1103202). I. A. T. and A. G. I. thank the Ministry of Science and Higher Education of the Russian Federation, the FSRC "Crystallography and Photonics" of RAS and the Russian Science Foundation (Project No. 19-12-00414) for support of this work. A.R.O. thanks the Russian Science Foundation (grant no. 19-72-30043).

A.G.K. and D.V.S. thank the Russian Foundation for Basic Research, grant no. 19-03-00100. D.V.S. thank the Russian Foundation for Basic Research, grants no. 20-32-90099. A.R.O and D.V.S. thank the Ministry of Science and Higher Education agreement No. 075-15-2020-808. N.V.T. and A.I.B. also thank the USA National Science Foundation (Grant CHE-1664379). A.G.G. acknowledge support from the Center for Collective Use Accelerator Center for Neutron Research of the Structure of Substance and Nuclear Medicine of the INR RAS. We thank Igor Grishin (Skoltech) for proofreading of the manuscript.

**Supporting Information Available:** Detailed description of used experimental and computational methods, structural information of predicted and synthesized structures, additional data on the Le Bail refinements, information about magnetic structure, natural population analysis, electronic structure, phonons, anharmonic calculations and elastic properties.

**Author Contributions**
Dmitrii V. Semenok, Di Zhou, and Alexander G. Kvashnin contributed equally to this work.
X.H. and A.R.O. conceived this project. D.Z., D.V.S., I.A.T, A.G.I., A.G.G. and X.H. performed the experiment. D.V.S., D.Z., A.R.O., and T.C. prepared the theoretical calculations and analysis, M.G. studied magnetic properties. X.H., D.Z., D.V.S., M.G., A.R.O., and T.C. wrote and revised the paper. All authors discussed the results and offered useful ideas.

**Competing Interests**
The authors declare no competing interests.

**Data and Materials Availability**
All data needed to evaluate the conclusions are presented in the paper and/or Supporting Information. Additional data related to this paper may be requested from the authors.

# Supporting Information

# Novel Strongly Correlated Europium Superhydrides

Dmitrii V. Semenok,[1] Di Zhou,[2] Alexander G. Kvashnin,[1] Xiaoli Huang,[2,*] Michele Galasso,[1] Ivan A. Kruglov,[3,4] Anna G. Ivanova,[5] Alexander G. Gavrilyuk,[5,6] Wuhao Chen,[2] Nikolay V. Tkachenko,[7] Alexander I. Boldyrev,[7] Ivan Troyan,[5] Artem R. Oganov,[1,*] and Tian Cui[8,2]

[1] Skolkovo Institute of Science and Technology, Skolkovo Innovation Center, 3 Nobel Street, Moscow 143026, Russia
[2] State Key Laboratory of Superhard Materials, College of Physics, Jilin University, Changchun 130012, China
[3] Moscow Institute of Physics and Technology, 9 Institutsky Lane, Dolgoprudny 141700, Russia
[4] Dukhov Research Institute of Automatics (VNIIA), Moscow 127055, Russia
[5] Shubnikov Institute of Crystallography, Federal Scientific Research Center Crystallography and Photonics, Russian Academy of Sciences, 59 Leninskii pr-t, Moscow 119333, Russia
[6] IC RAS Institute for Nuclear Research, Russian Academy of Sciences, Moscow 117312, Russia
[7] Department of Chemistry and Biochemistry, Utah State University, 0300 Old Main Hill, Logan, Utah, 84322-0300, USA
[8] School of Physical Science and Technology, Ningbo University, Ningbo 315211, China

**Corresponding Authors**

Dr. X. Huang, e-mail: huangxiaoli@jlu.edu.cn, Prof. A. Oganov, e-mail: a.oganov@skoltech.ru

# Contents

# Methods

*Experiment*

To investigate the formation of new chemical compounds in the Eu–H system, we purchased the europium foil with 99.99% purity from Alfa Aesar company. Two diamond anvil cells (DACs) with 100 and 150 µm culets were loaded with Eu and ammonia borane (AB), purified by sublimation, which was used as a source of hydrogen and a pressure transmitting medium (Table S1). As has been shown earlier in the synthesis of superhydrides of lanthanum,[1] thorium,[2] praseodymium,[3] and neodymium,[4] ammonia borane is an effective source of hydrogen when the metal target is heated by a short (< 0.2 s) laser pulse due to the well-known thermal decomposition reaction: $NH_3BH_3 \rightarrow H_2 + poly\text{-}(BNH_x)_n$.[5,6]

A tungsten plate with a thickness of $20 \pm 2$ µm was used as a gasket. The pressure was determined by the Raman signal of diamond.[7] Heating was carried out by pulses of an infrared laser (1 µm, Nd:YAG). A part of the X-ray diffraction (XRD) patterns of all samples studied in diamond anvil cells were recorded on 4W2 beamline of the Beijing Synchrotron Research Facilities (BSRF, China) and BL15U1 synchrotron beamline at the Shanghai Synchrotron Research Facility (SSRF, China)[8] using a focused (5×12 µm) monochromatic X-ray beam (20 keV, 0.6199 Å). The other part of the high-pressure XRD patterns were obtained at BL10XU in SPring-8 using monochromatic synchrotron radiation and an imaging plate detector at room temperature.[9] The X-ray beam with a wavelength of 0.413 Å was focused with a polymer compound refractive lens (SU-8, produced by ANKA). The experimental X-ray diffraction images were analyzed and integrated using Dioptas software package.[10] The full profile analysis of the diffraction patterns and calculations of the unit cell parameters were performed in JANA2006[11] using the Le Bail method.[12] Additional parameters of high-pressure diamond anvil cells are listed in Table **S1**.

**Table S1.** Experimental parameters of the DACs used to synthesize europium hydrides.

| Cell | Synthesis pressure, GPa | Culet size, µm | Sample size, µm | Composition/load |
|------|------------------------|----------------|-----------------|------------------|
| E1   | 110                    | 100            | 15              | $Eu/BH_3NH_3$    |
| E2   | 74                     | 150            | 20              | $Eu/BH_3NH_3$    |
| E3   | 86                     | 100            | 13              | $Eu/BH_3NH_3$    |

*Theory*

All calculations of crystalline systems were carried out using the Vienna Ab initio Simulation Package (VASP) code[13–15] with PAW[16–18] pseudopotentials. The calculations were performed solely for producing the wave functions of the systems. The generalized gradient approximation (GGA) expressed by the PBE functional was applied. The Brillouin zone was sampled using the Monkhorst–Pack method.[19] Considering the different number of atoms in unit cells for different compounds, various $k$-grids (5×5×5 and 11×11×11) were used. The energy cutoff was set to 600 eV, the partial occupancies were set using the Gaussian smearing with $\sigma = 0.1$. To find the bonding pattern of the selected clusters, the solid-state adaptive natural density partitioning (SSAdNDP)[20] algorithm was implemented. The SSAdNDP follows the idea of the AdNDP algorithm for molecules[21,22] and the periodic NBO method.[23] It is based on the partitioning of the electron density (1-body density matrix) and produces the interpretation of chemical bonding in systems with translational symmetry in terms of classical lone pairs and two-center bonds, as well as multicenter delocalized bonding. The algorithm produces the number of bonding elements and their occupancy numbers (ONs). Physically, it shows how many electrons are sitting in the chosen region. For the ideal Lewis case, the ON is 2.00 |$e$|. It has been shown that the chemical bonding pattern obtained with the SSAdNDP has a high correlation with features of studied materials and can give important insights into the physical and chemical properties of various solids.[24–28] The Stuttgart RSC 1997 ECP atomic centered basis[29] for actinides and Def2-TZVPP[30] basis for other atoms were used to represent the projected PW density. For convenience, we designate this basis set combination as Basis_1. The spillage parameter of occupied bands for this basis was no higher than 1%. For spin-polarized calculations, both spin-up (alpha) and spin-down (beta) density matrices were analyzed. The resulting bonding elements were designated as alpha and beta bonds. The occupancies of doubly occupied bonding elements for spin-polarized cases were calculated as a sum of the occupancies of alpha and beta bonding elements.

For calculations of the model clusters, the Gaussian 16 program[31] was used. The exact geometries of the solid-state hydrides were taken for modeling the wave functions. The charges of clusters were set so as to take into account the stoichiometry of the crystals. All calculations were carried out at the PBE0/Basis_1 level of theory. The ChemCraft 1.8 software[32] was used to visualize the chemical bonding patterns and geometries of the investigated hydrides.

The calculations of the electron–phonon coupling and superconducting $T_C$ were carried out with QUANTUM ESPRESSO (QE) package[32] using the density functional perturbation theory,[33] employing the plane-wave generalized gradient approximation with the Goedecker–Hartwigsen–Hutter–Teter pseudopotentials.[34,35] In our ab initio calculations of the electron–phonon coupling (EPC) parameter λ of EuH$_9$, the first Brillouin zone was sampled by 2×2×2 $q$-points mesh and 10×10×10 $k$-points meshes with a smearing σ = 0.005–0.05 Ry that approximates the zero-width limits in the calculation of λ. The critical temperature $T_C$ was calculated using the Allen–Dynes equations.[36]

To calculate $U$–$J$, we took the most stable collinear states for each of the three phases (AFM 2 $F\bar{4}3m$-EuH$_9$, FM $P6_3/mmc$-EuH$_9$, and FM $Pm\bar{3}n$-Eu$_8$H$_{46}$) and the computed DFT ground state with $U$–$J$ = 0. Then, we computed the "bare" and "interacting" responses by applying a series of small perturbations (α = –0.08, –0.05, –0.02, 0.02, 0.05, 0.08) to one of the Eu sites. We obtained the response functions from a linear fit of the number of f electrons on the perturbed atomic site as a function of the applied potential α (Figure S4). From the slopes of these functions we obtained the value of $U$–$J$ (Table S6). The obtained results were close to those previously described in the study of $P6_3/mmc$-NdH$_9$.[4]

Before starting the magnetic calculations, we checked the parameters for convergence. After several tests with different VASP pseudopotentials (Eu_2, Eu), we noticed that the computed magnetization on each atom was always zero, which is physically meaningless. The only pseudopotential which allowed us to study the magnetic properties of EuH$_9$ is Eu (PAW_PBE_Eu_23Dec2003, $z_{val}$ = 17), so we used it together with that of H, both taken from the VASP library. The kinetic energy cutoff of the plane wave basis set was chosen to be 540 eV (520 eV) for the $F\bar{4}3m$ ($P6_3/mmc$) phase, which gives a maximum error of 1 meV/atom with respect to more accurate calculations. The same error criterion was used when choosing, for both EuH$_9$ phases, a length $l$ = 40 for the automatic generation of GAMMA-centered Monkhorst–Pack grids as implemented in the VASP code, and a smearing parameter SIGMA = 0.2 with the Methfessel–Paxton method of order 1.

# Structural Information

**Table S2.** Calculated (SOC, DFT+U, $U$–$J$ = 5 eV) crystal structure of the discovered Eu–H phases.

| Phase | Pressure, | Lattice | Coordinates | | | |
|---|---|---|---|---|---|---|
| $P6_3/mmc$-EuH$_9$ | 100 | $a$ = 3.559 Å<br>$c$ = 5.855 Å<br>$\alpha = \beta = \gamma = 90°$ | Eu1 | 0.33333 | 0.66667 | 0.75000 |
| | | | H1 | 0.00000 | 0.00000 | 0.25000 |
| | | | H2 | 0.33333 | 0.66667 | 0.33627 |
| | | | H3 | 0.15200 | 0.30401 | 0.04717 |
| $F\bar{4}3m$-EuH$_9$ | 100 | $a$ = 5.041 Å<br>$\alpha = \beta = \gamma = 90°$ | Eu1 | 0.25000 | 0.25000 | 0.25000 |
| | | | H1 | -0.13655 | -0.13655 | -0.13655 |
| | | | H2 | -0.38097 | -0.38097 | -0.38097 |
| | | | H3 | 0.00000 | 0.00000 | 0.00000 |
| $Pm\bar{3}n$-Eu$_4$H$_{46}$ | 130 | $a$ = 5.865 Å<br>$\alpha = \beta = \gamma = 90°$ | Eu1 | 0.00000 | 0.00000 | 0.00000 |
| | | | Eu2 | 0.25000 | 0.00000 | 0.50000 |
| | | | H1 | 0.00000 | 0.11963 | 0.30584 |
| | | | H2 | 0.25000 | 0.50000 | 0.00000 |
| | | | H3 | 0.18451 | 0.18451 | 0.18451 |
| $Fm\bar{3}n$-EuH$_{10}$ | 100 | $a$ = 5.183 Å<br>$\alpha = \beta = \gamma = 90°$ | Eu | 0.00000 | 0.00000 | 0.00000 |
| | | | H1 | 0.11916 | 0.11916 | 0.11916 |
| | | | H2 | 0.25000 | 0.25000 | 0.25000 |
| $P1$-EuH$_9$ (pseudohexagonal) | 130 | $a$=3.5598 Å<br>$b$=6.8524 Å<br>$c$=6.1657 Å<br>$\alpha$=63.26°<br>$\beta$=90°<br>$\gamma$=74.94° | Eu1 | 0.375 | 0.25 | 0.708333 |
| | | | Eu2 | 0.625 | 0.75 | 0.291667 |
| | | | Eu3 | 0.875 | 0.25 | 0.208333 |
| | | | Eu4 | 0.125 | 0.75 | 0.791667 |
| | | | H1 | 0.625 | 0.75 | 0.625 |
| | | | H2 | 0.125 | 0.75 | 0.125 |
| | | | H3 | 0.875 | 0.25 | 0.875 |
| | | | H4 | 0.375 | 0.25 | 0.375 |
| | | | H5 | 0.918135 | 0.16373 | 0.584801 |
| | | | H6 | 0.418135 | 0.16373 | 0.084801 |
| | | | H7 | 0.081865 | 0.83627 | 0.415199 |
| | | | H8 | 0.581865 | 0.83627 | 0.915199 |
| | | | H9 | 0.168135 | 0.66373 | 0.501468 |
| | | | H10 | 0.668135 | 0.66373 | 0.001468 |
| | | | H11 | 0.831865 | 0.33627 | 0.498532 |
| | | | H12 | 0.331865 | 0.33627 | 0.998532 |
| | | | H13 | 0.501592 | 0.452831 | 0.349587 |
| | | | H14 | 0.001592 | 0.452831 | 0.849587 |
| | | | H15 | 0.498408 | 0.547169 | 0.650413 |
| | | | H16 | 0.998408 | 0.547169 | 0.150413 |
| | | | H17 | 0.773584 | 0.452831 | 0.621579 |
| | | | H18 | 0.273584 | 0.452831 | 0.121579 |
| | | | H19 | 0.226416 | 0.547169 | 0.378421 |
| | | | H20 | 0.726416 | 0.547169 | 0.878421 |
| | | | H21 | 0.045577 | 0.452831 | 0.349587 |
| | | | H22 | 0.545577 | 0.452831 | 0.849587 |
| | | | H23 | 0.954423 | 0.547169 | 0.650413 |

| | | | H24 | 0.454423 | 0.547169 | 0.150413 |
| | | | H25 | 0.295577 | 0.952831 | 0.447582 |
| | | | H26 | 0.795577 | 0.952831 | 0.947582 |
| | | | H27 | 0.704423 | 0.047169 | 0.552418 |
| | | | H28 | 0.204423 | 0.047169 | 0.052418 |
| | | | H29 | 0.023584 | 0.952831 | 0.175589 |
| | | | H30 | 0.523584 | 0.952831 | 0.675589 |
| | | | H31 | 0.976416 | 0.047169 | 0.82441 |
| | | | H32 | 0.476416 | 0.047169 | 0.32441 |
| | | | H33 | 0.751592 | 0.952831 | 0.447582 |
| | | | H34 | 0.251592 | 0.952831 | 0.947582 |
| | | | H35 | 0.248408 | 0.047169 | 0.552418 |
| | | | H36 | 0.748408 | 0.047169 | 0.052418 |
| *P*1-EuH₉ (pseudocubic) | 130 | *a*=3.5646 Å *b*=3.5646 Å *c*=6.1742 Å *α*=90° *β*=73.22° *γ*=60° | Eu1 | 0.75 | 0.125 | 0.375 |
| | | | Eu2 | 0.75 | 0.625 | 0.875 |
| | | | H1 | 0.59035 | 0.204825 | 0.068275 |
| | | | H2 | 0.59035 | 0.704825 | 0.568275 |
| | | | H3 | 0.13655 | 0.658625 | 0.068275 |
| | | | H4 | 0.13655 | 0.158625 | 0.568275 |
| | | | H5 | 0.13655 | 0.431725 | 0.295175 |
| | | | H6 | 0.13655 | 0.931725 | 0.795175 |
| | | | H7 | 0.13655 | 0.204825 | 0.068275 |
| | | | H8 | 0.13655 | 0.704825 | 0.568275 |
| | | | H9 | 0.857095 | 0.571452 | 0.190484 |
| | | | H10 | 0.857095 | 0.071452 | 0.690484 |
| | | | H11 | 0.380968 | 0.047579 | 0.190484 |
| | | | H12 | 0.380968 | 0.547579 | 0.690484 |
| | | | H13 | 0.380968 | 0.809515 | 0.428548 |
| | | | H14 | 0.380968 | 0.309515 | 0.928548 |
| | | | H15 | 0.380968 | 0.571452 | 0.190484 |
| | | | H16 | 0.380968 | 0.071452 | 0.690484 |
| | | | H17 | 0.0 | 0.0 | 0.0 |
| | | | H18 | 0.0 | 0.5 | 0.5 |

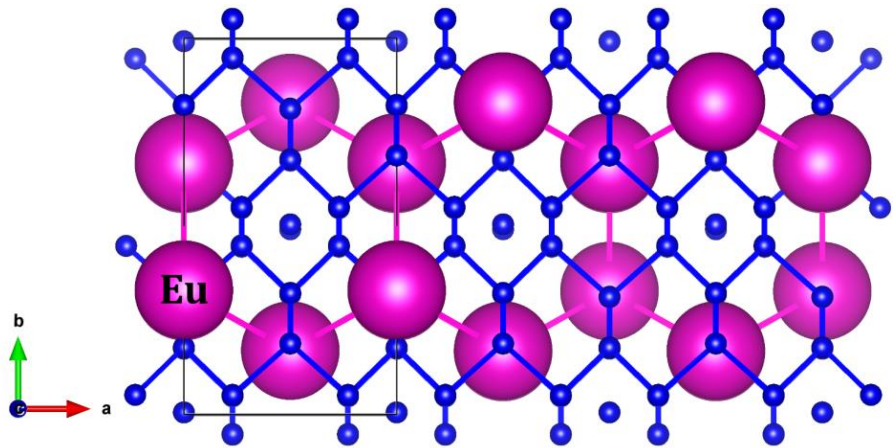

**Figure S1.** Crystal structure of pseudohexagonal *P*1(hex)-EuH₉ (=Eu₄H₃₆) with the disordered hydrogen sublattice, visualized using VESTA software.[37] The hydrogen atoms are shown in blue.

**Table S3.** Experimental and theoretical cell parameters of all studied Eu–H phases. $V_{DFT+U}$ is calculated with SOC and $U–J = 5$ eV, $V_{DFT}$ is calculated without the Hubbard correction. Data in brackets correspond to experiment at $\lambda = 0.6199$ Å and 0.413 Å.

| Compound | Pressure, GPa | $a$, Å | $c$, Å | $V$, Å³ per Eu atom | $V_{DFT+U}$, Å³ per Eu atom | $V_{DFT}$, Å³ per Eu atom |
|---|---|---|---|---|---|---|
| $Pm\bar{3}n$-Eu$_8$H$_{46}$ ($Z = 8$) | 130 | 5.8582 | | 25.13 | 25.21 | 24.75 |
| | 125 | 5.8604 | | 25.16 | 25.49 | 25.02 |
| | 121 | 5.8649 | | 25.22 | 25.77 | 25.29 |
| | 116 | 5.8862 | | 25.49 | 26.08 | 25.58 |
| | 111 | 5.9069 (5.9080) | | 25.76 (25.77) | 26.38 | 25.88 |
| | 106 | 5.9492 | | 26.32 | 26.65 | 26.15 |
| | 102 | 5.9547 | | 26.39 | 26.92 | 26.38 |
| | 99 | 5.9656 | | 26.54 | 27.13 | 26.58 |
| | 97 | 5.9729 | | 26.63 | 27.26 | 26.71 |
| | 94 | 5.9732 | | 26.64 | 27.54 | 26.96 |
| | 90 | 5.9839 | | 26.78 | 27.79 | 27.21 |
| | 89 | 6.0125 | | 27.17 | 27.80 | 26.87 |
| | 86 | 5.9973 | | 26.96 | 28.12 | 27.51 |
| $P6_3/mmc$-EuH$_9$ ($Z = 2$) | 130 | 3.5911 (3.5887) | 5.5094 (5.5027) | 30.76 (30.68) | 30.01 | 29.34 |
| | 125 | 3.5870 | 5.5176 | 30.74 | 31.38 | 30.53 |
| | 121 | 3.5905 | 5.5265 | 30.85 | 31.67 | 30.85 |
| | 118 | 3.5992 | 5.5419 | (31.085) | 31.89 | 31.00 |
| | 116 | 3.5914 | 5.6063 | 31.31 | 32.07 | 31.19 |
| | 111 | 3.6203 | 5.5171 | 31.31 | 32.51 | 31.62 |
| $F\bar{4}3m$-EuH$_9$ ($Z = 4$) | 130 | 4.9475 | | 30.27 | 30.67 | 29.92 |
| | 125 | 4.9472 | | 30.27 | 30.98 | 30.28 |
| | 121 | 4.9566 | | 30.44 | 31.26 | 30.57 |
| | 116 | 4.9741 | | 30.77 | 31.63 | 30.90 |
| | 111 | 4.9786 | | 30.85 | 32.04 | 31.33 |
| | 106 | 5.0313 | | 31.84 | 32.43 | 31.67 |
| | 102 | 5.0439 | | 32.08 | 32.76 | 32.00 |
| | 99 | 5.0589 | | 32.37 | 33.01 | 32.23 |
| | 97 | 5.0700 | | 32.58 | 33.20 | 32.42 |
| | 94 | 5.0765 | | 32.71 | 33.48 | 32.71 |
| | 90 | 5.0866 | | 32.90 | 33.82 | 32.93 |
| | 86 | 5.1088 | | 33.33 | 34.15 | 33.39 |

**Table S4.** Calculated EoS parameters in the 3rd order Birch–Murnaghan equation with $K_0'$ fixed 4 for $Pm\bar{3}n$-Eu$_8$H$_{46}$, $P6_3/mmc$-EuH$_9$, and $F\bar{4}3m$-EuH$_9$. $V_0$ for all europium hydrides corresponds to 100 GPa.

|  | $Pm\bar{3}n$-Eu$_8$H$_{46}$ | $P6_3/mmc$-EuH$_9$ | $F\bar{4}3m$-EuH$_9$ |
|---|---|---|---|
| $V_0$ (Å$^3$) | 26.3(1) | 31.9(1) | 31.3(1) |
| $K_0$ (GPa) | 471 ± 70 | 594 ± 70 | 699 ± 27 |
| $K_0'$ | 4 | 4 | 4 |

**Table S5.** Relative energies (in kcal/mol) of the Eu clusters at different spin states.

| Spin state ($S_{total}$) | EuH$_{28}$$^{19+}$ | EuH$_{27}$$^{18+}$ | EuH$_{24}$$^{17+}$ |
|---|---|---|---|
| 0 | 207.78 | 235.09 | 211.30 |
| 1 | 91.34 | 100.62 | 112.89 |
| 2 | 80.32 | 84.52 | 84.40 |
| 3 | 0 | 9.73 | 64.18 |
| 4 | 9.62 | 0 | 0 |

**Table S6.** Occupancies of valence superatomic orbitals.

| MO Type | Hexagonal EuH$_9$ Alpha (↑) | Hexagonal EuH$_9$ Beta (↓) | MO Type | Cubic EuH$_9$ Alpha (↑) | Cubic EuH$_9$ Beta (↓) | MO type | $Pm\bar{3}n$-Eu$_8$H$_{46}$ Alpha (↑) | $Pm\bar{3}n$-Eu$_8$H$_{46}$ Beta (↓) |
|---|---|---|---|---|---|---|---|---|
| F | 0.97 | - | F | 0.97 | - | F | 0.98 | - |
| F | 0.97 | - | F | 0.97 | - | F | 0.98 | - |
| F | 0.97 | - | F | 0.97 | - | F | 0.98 | - |
| F | 0.97 | - | F | 0.97 | - | F | 0.98 | - |
| F | 0.97 | - | F | 0.96 | - | F | 0.97 | - |
| F | 0.97 | - | F | 0.96 | - | F | 0.96 | - |
| F | 0.96 | - | F | 0.96 | - | F | 0.96 | - |
| D | 0.95 | 0.77 | D | 0.79 | 0.78 | D | 0.59 | 0.58 |
| D | 0.78 | 0.76 | D | 0.79 | 0.77 | D | 0.59 | 0.58 |
| D | 0.77 | 0.74 | P | 0.68 | 0.70 | D | 0.59 | 0.56 |
| D | 0.74 | 0.72 | P | 0.68 | 0.70 | D | 0.56 | 0.55 |
| D | 0.72 | 0.67 | P | 0.68 | 0.70 | D | 0.55 | 0.54 |
| P | - | 0.65 | S | - | 0.65 | P | - | 0.54 |



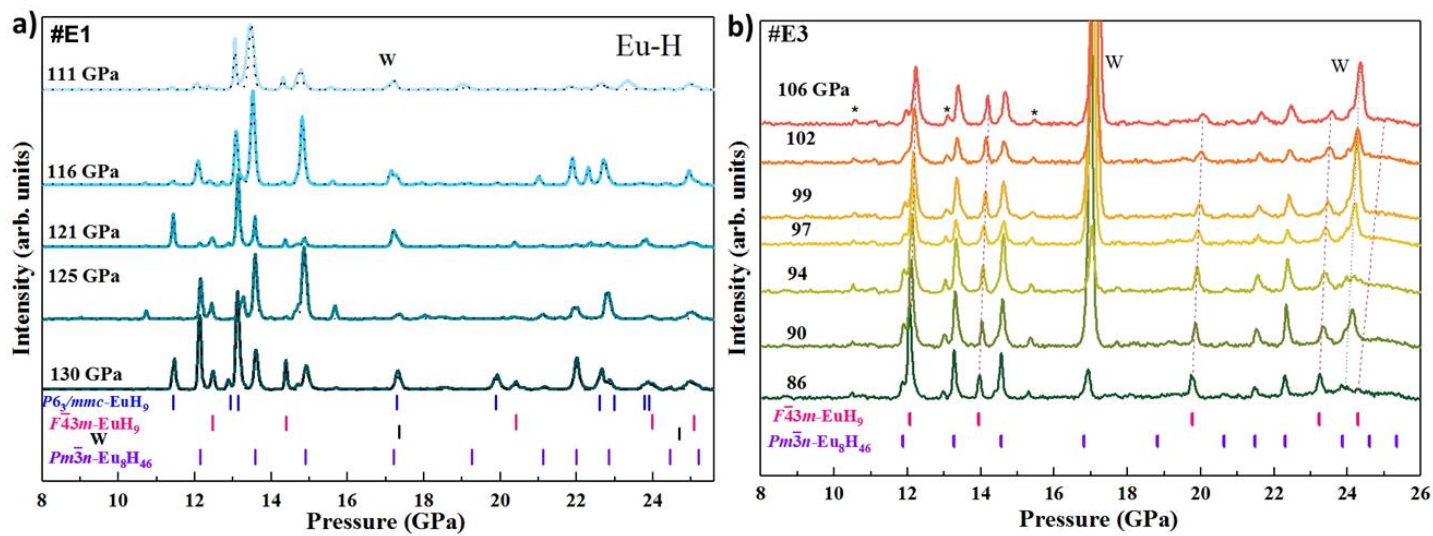

**Figure S2.** Decomposition of Eu polyhydrides after decreasing pressure in cell **(a)** #E1 and **(b)** #E3. The XRD patterns were indexed by $Pm\bar{3}n$-$Eu_8H_{46}$, $P6_3/mmc$-$EuH_9$, and $F\bar{4}3m$-$EuH_9$. Dash lines are the model fit for the structures in Figures S2a.

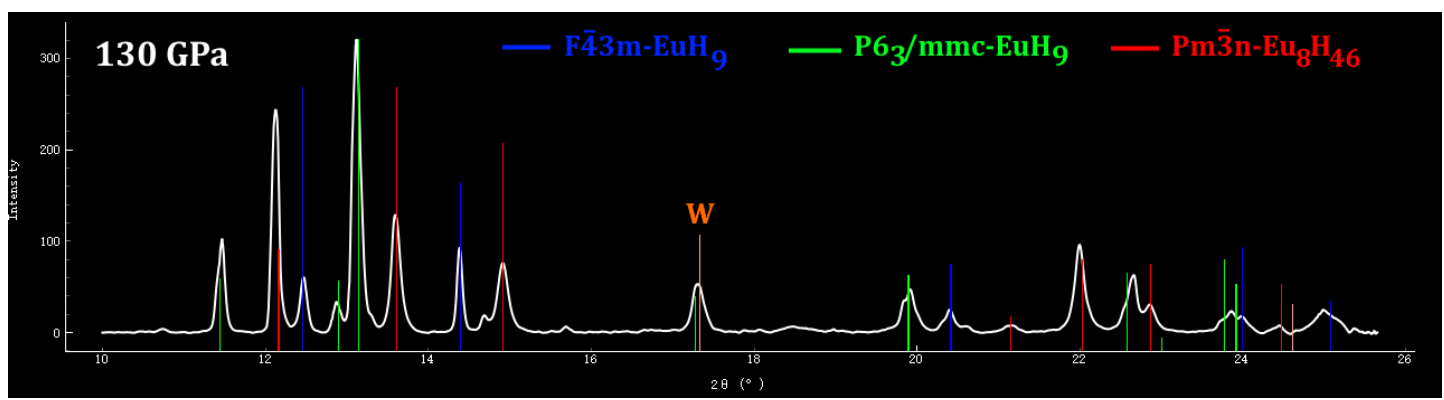

**Figure S3.** Qualitative correlation of the experimental reflections and simulated diffraction patterns of the Eu–H phases at 130 GPa, integrated using Dioptas software.

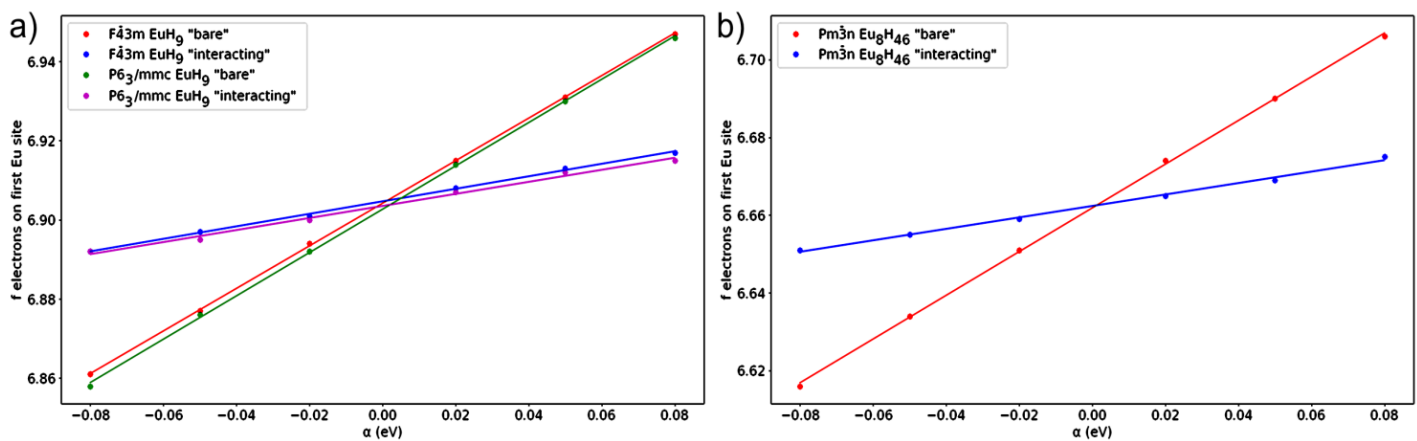

**Figure S4.** Linear response functions for (a) both $EuH_9$ phases and (b) the $Eu_8H_{46}$ phase. The calculated values are shown by dots; straight lines represent the least squares fit.



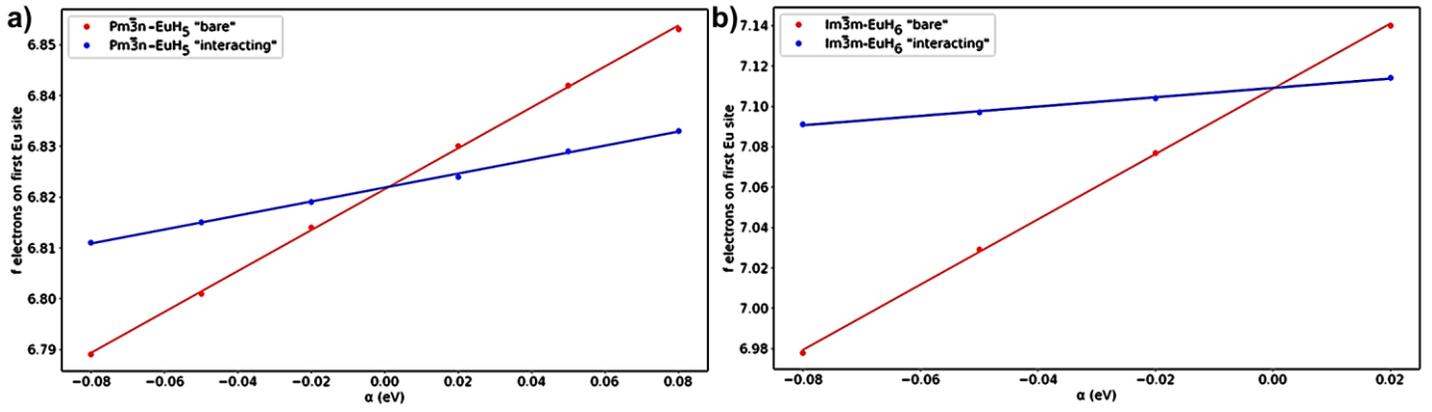

**Figure S5.** Linear response functions for (a) EuH$_5$ and (b) EuH$_6$ phases. The calculated values are shown by dots; straight lines represent the least squares fit.

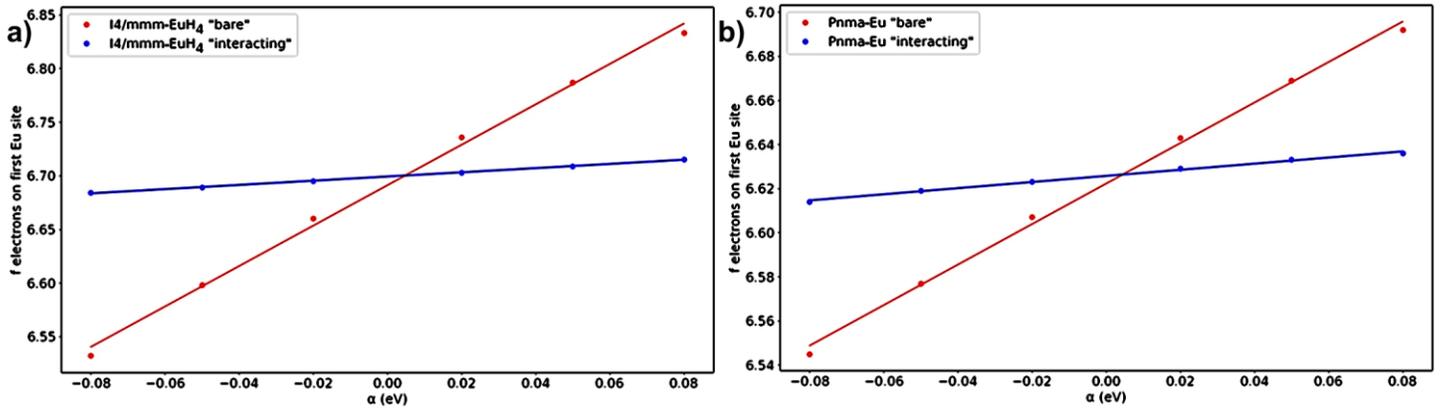

**Figure S6.** Linear response functions for (a) tetragonal EuH$_4$ and (b) pure Eu at 75 GPa. The calculated values are shown by dots; straight lines represent the least squares fit.

**Table S7.** Linear response functions and calculated $U$ for the Eu–H phases at 130 GPa.

| Phase | "Bare" response (eV$^{-1}$) | "Interacting" response (eV$^{-1}$) | $U$–$J$ (eV) |
|---|---|---|---|
| $F\bar{4}3m$-EuH$_9$ | 0.158 | 0.537 | 4.46 |
| $P6_3/mmc$-EuH$_9$ | 0.152 | 0.547 | 4.74 |
| $Pm\bar{3}n$-Eu$_8$H$_{46}$ | 0.147 | 0.562 | 5.01 |
| $Im\bar{3}m$-EuH$_6$ * | 0.231 | 1.616 | 3.7 |
| $Pm\bar{3}n$-EuH$_5$ | 0.138 | 0.403 | 4.8 |
| $I4/mmm$-EuH$_4$ | 0.196 | 1.884 | 4.6 |
| $Pnma$-Eu (75 GPa) | 0.139 | 0.918 | 6.1 |

*For EuH$_6$, only perturbations up to 0.02 eV were included, because larger perturbations showed high nonlinearity.



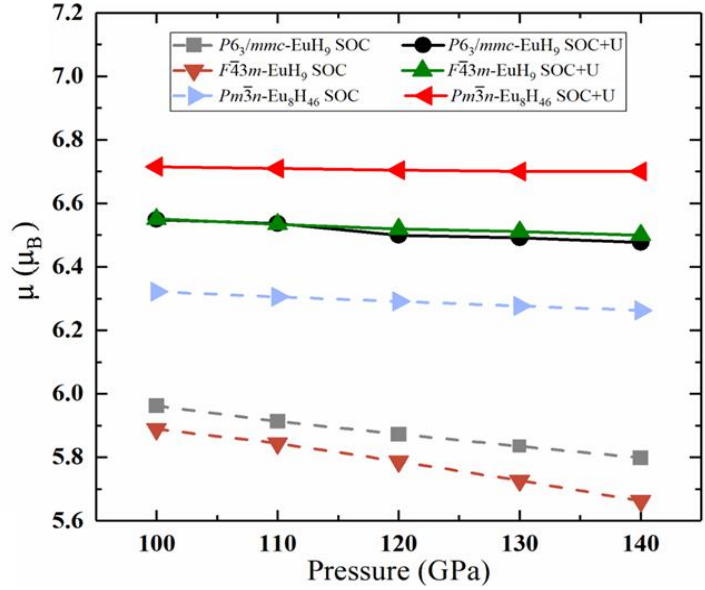

**Figure S7.** Magnetic moments (μ$_B$ per 1 Eu atom) of all synthesized Eu-H compounds with and without $U$–$J$ and SOC.

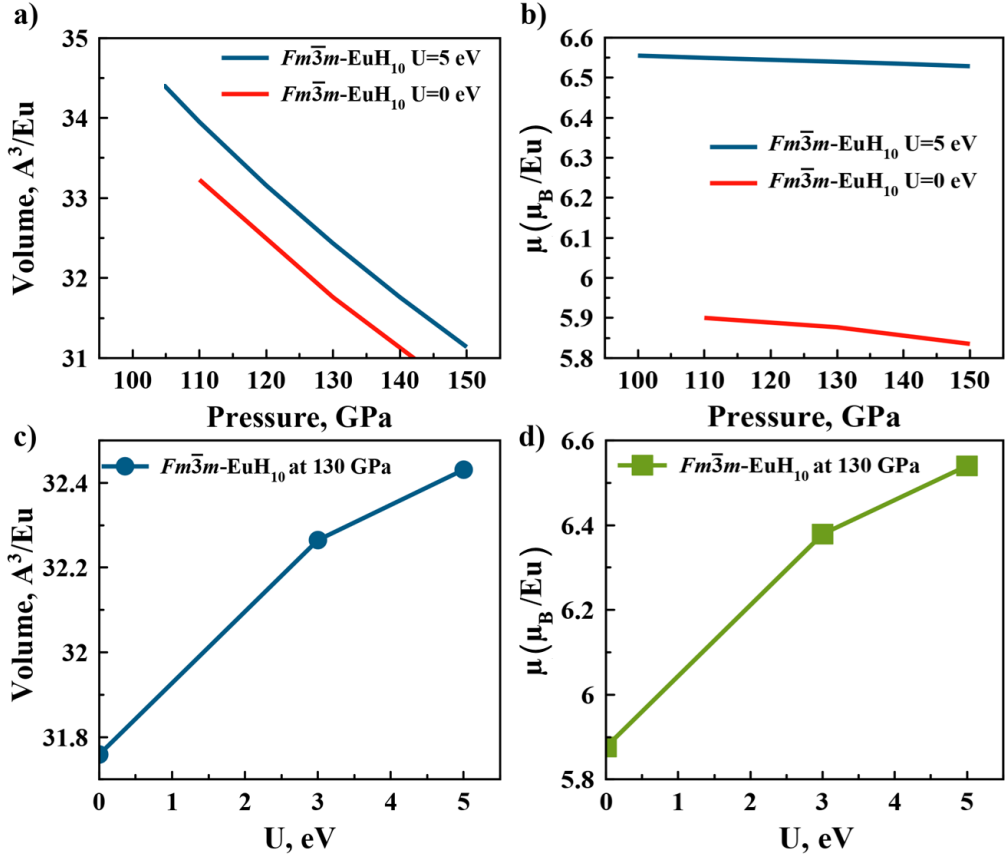

**Figure S8.** (a) Equation of state, (b) magnetic moment per 1 Eu atom, (c) dependence of the unit cell volume $V$ on the Hubbard-like correction term $U$–$J$ at 130 GPa, and (d) dependence of the magnetic moment on $U$–$J$ at 130 GPa for the proposed $Fm\bar{3}m$-EuH$_{10}$, isostructural to $Fm\bar{3}m$-LaH$_{10}$.[38]



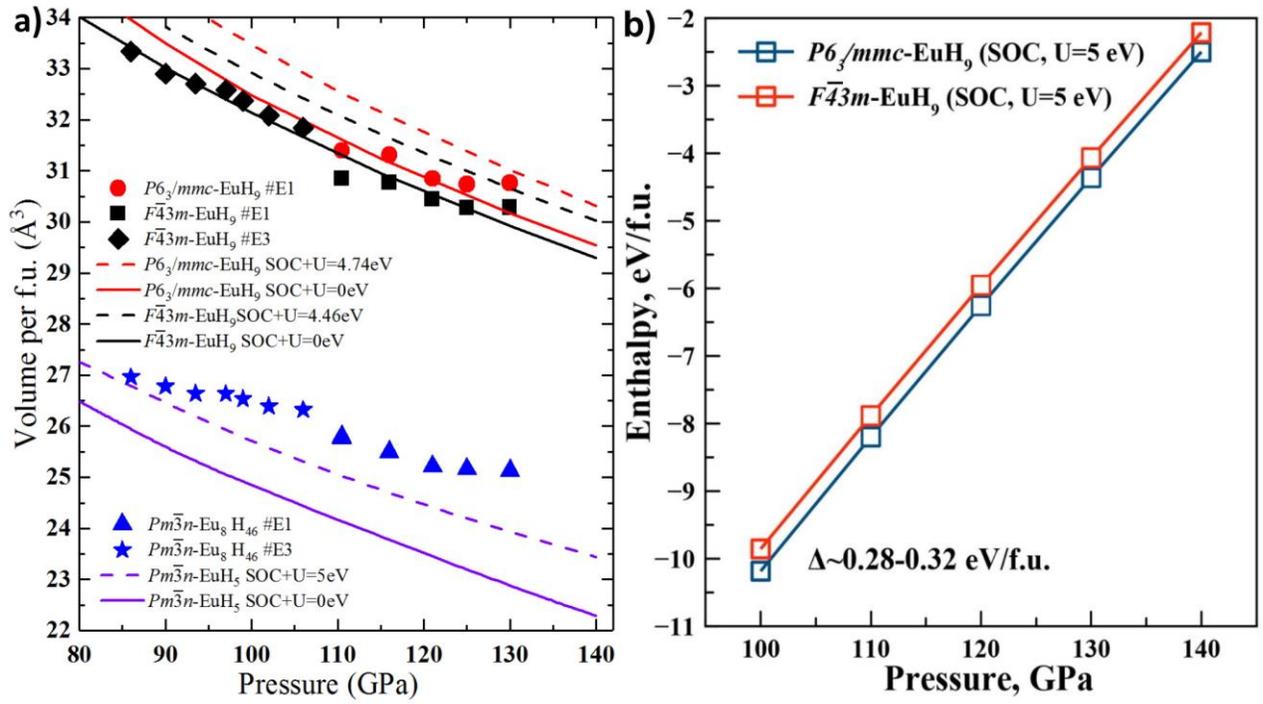

**Figure S9.** (a) Experimental cell volumes of the discovered Eu hydrides and calculated equations of states for $F\bar{4}3m$-EuH$_9$, $P6_3/mmc$-EuH$_9$, and proposed $Pm\bar{3}n$-Eu$_8$H$_{46}$[19] with SOC and $U$–$J$. (b) Difference between enthalpies of cubic and hexagonal modifications of EuH$_9$.

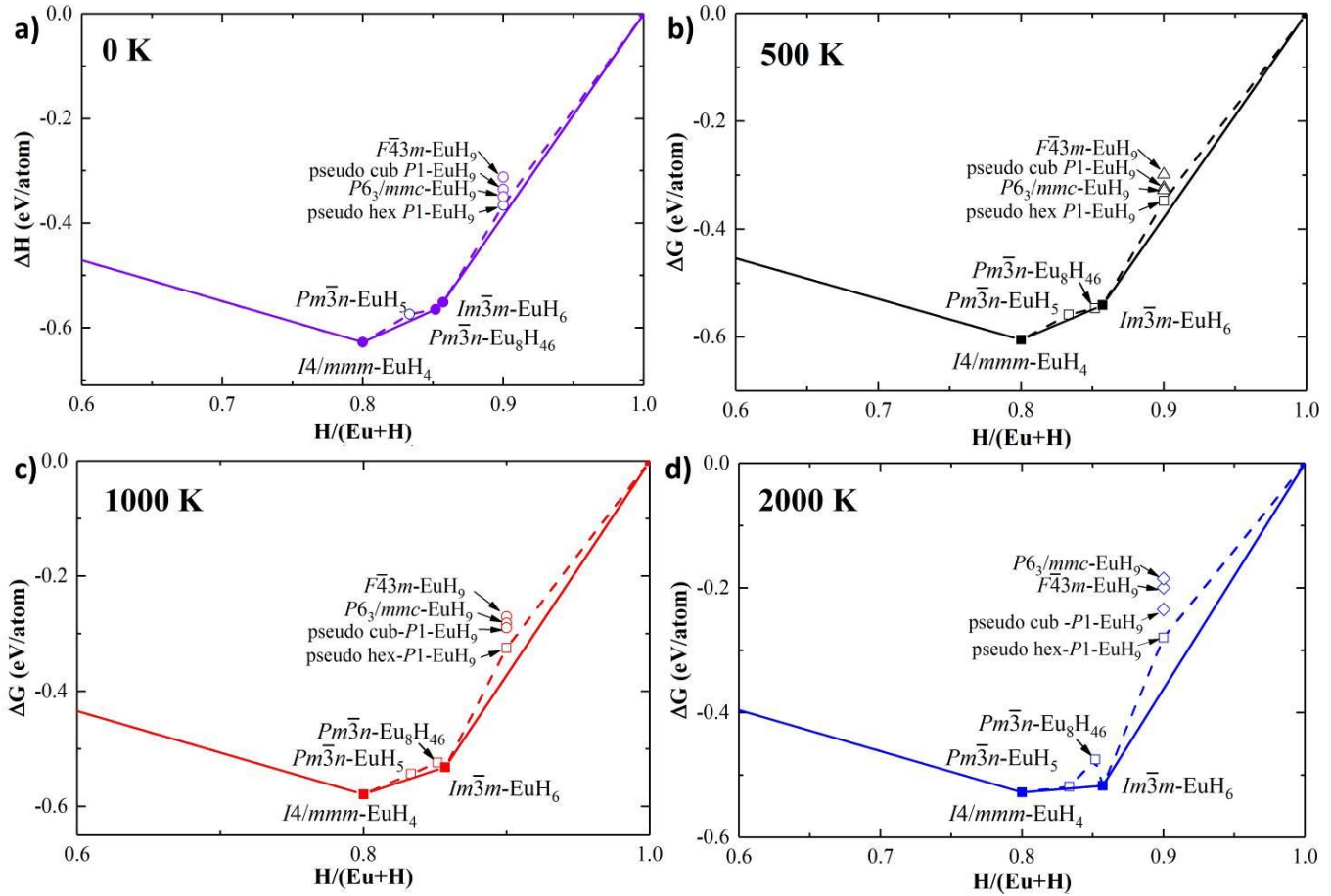

**Figure S10.** Calculated convex hulls of the Eu–H system at 130 GPa and (a) 0 K, (b) 500 K, (c) 1000 K, and (d) 2000 K with the the zero-point energy (ZPE), SOC, and $U$–$J$, specific for each compound (see Table S8).



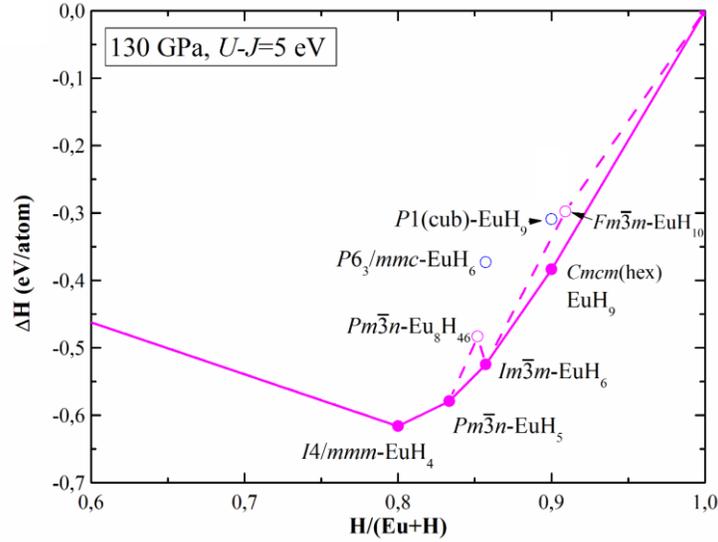

**Figure S11.** Calculated convex hulls of the Eu–H system at 130 GPa for 0 K with the ZPE, SOC, and fixed $U$–$J$ = 5 eV.

**Table S8.** Enthalpy and energy of formation with and without the ZPE for various Eu–H phases at 130 GPa with SOC and $U$–$J$ (from Table S7).

| Phase | Eu | H | $x$ (H/Eu+H) | ZPE, eV/Eu | E+ZPE, eV/Eu | $H_{form}$, eV/Eu |
|---|---|---|---|---|---|---|
| $Pnma$-Eu | 4 | 0 | 0 | 0.0047 | 5.42868 | 0 |
| $I4/mmm$-EuH$_4$ | 2 | 8 | 0.8 | 0.15492 | –0.86015 | –0.62783 |
| $Pm\bar{3}n$-EuH$_5$ | 8 | 40 | 0.83333 | 0.19861 | –0.85113 | –0.57414 |
| $Pm\bar{3}n$-Eu$_8$H$_{46}$ | 8 | 46 | 0.85185 | 0.20604 | –1.94194 | –0.56507 |
| $Im\bar{3}m$-EuH$_6$ | 2 | 12 | 0.85714 | 0.21108 | –2.90023 | –0.5512 |
| $F\bar{4}3m$-EuH$_9$ | 4 | 36 | 0.9 | 0.1459 | –3.14065 | –0.31209 |
| $P6_3/mmc$-EuH$_9$ | 2 | 18 | 0.9 | 0.19861 | –4.75865 | –0.34938 |
| pseudocubic EuH$_9$ | 2 | 1 | 0.9 | 0.34406 | –5.13153 | –0.33561 |
| pseudohexagonal EuH$_9$ | 4 | 1 | 0.9 | 0.35587 | –4.99386 | –0.36576 |
| $C2/c$-H | 0 | 24 | 1 | 0.00174 | –0.78516 | 0 |

**Table S9.** Temperature dependence of the Gibbs free energy of formation ($G_{form}$, eV/atom), computed with the ZPE, SOC, and $U$–$J$ (from Table S7) for various Eu–H phases at 130 GPa.

| Phase \ Temperature, K | 0 | 500 | 1000 | 2000 |
|---|---|---|---|---|
| $Pnma$-Eu | 0 | 0 | 0 | 0 |
| $I4/mmm$-EuH$_4$ | –0.62783 | –0.60517 | –0.57896 | –0.52767 |
| $Pm\bar{3}n$-EuH$_5$ | –0.57414 | –0.55873 | –0.54337 | –0.51834 |
| $Pm\bar{3}n$-Eu$_8$H$_{46}$ | –0.56507 | –0.54654 | –0.5233 | –0.47524 |
| $Im\bar{3}m$-EuH$_6$ | –0.5512 | –0.54096 | –0.53178 | –0.51735 |
| $F\bar{4}3m$-EuH$_9$ | –0.31209 | –0.29836 | –0.2704 | –0.19966 |
| $P6_3/mmc$-EuH$_9$ | –0.34938 | –0.32308 | –0.28185 | –0.18473 |
| pseudocubic EuH$_9$ | –0.33561 | –0.31639 | –0.28987 | –0.23441 |
| pseudohexagonal EuH$_9$ | –0.36576 | –0.34746 | –0.32449 | –0.2797 |
| $C2/c$-H | 0 | 0 | 0 | 0 |



**Table S10.** Results of the fixed-composition USPEX search for the best crystal structure of $Pm\bar{3}n$-Eu$_8$H$_{46}$ at 130 GPa.

| ID | Origin | Enthalpy | Volume (A$^3$) | Density (g/cm$^3$) | Fitness (eV) |
|---|---|---|---|---|---|
| 242 | Permutate | –32.495 | 189.161 | 11.079 | –32.495 |
| 159 | keptBest | –32.494 | 189.161 | 11.079 | –32.494 |
| 70 | keptBest | –32.488 | 189.161 | 11.079 | –32.488 |
| 37 | Seeds | –32.485 | 189.161 | 11.079 | –32.485 |
| 114 | keptBest | –28.975 | 189.161 | 11.079 | –28.975 |
| 197 | keptBest | –28.474 | 189.161 | 11.079 | –28.474 |
| 247 | Permutate | –27.51 | 186.344 | 11.247 | –27.51 |
| 127 | keptBest | –27.458 | 185.838 | 11.277 | –27.458 |
| 8 | Random | –22.699 | 180.853 | 11.588 | –22.699 |
| 271 | softmutate | –22.578 | 181.419 | 11.552 | –22.578 |
| 295 | softmutate | –22.225 | 186.344 | 11.247 | –22.225 |
| 272 | softmutate | –22.049 | 181.419 | 11.552 | –22.049 |
| 42 | RandTop | –21.987 | 180.853 | 11.588 | –21.987 |
| 200 | Heredity | –21.948 | 184.714 | 11.346 | –21.948 |
| 199 | Heredity | –21.858 | 185.893 | 11.274 | –21.858 |
| 131 | softmutate | –21.418 | 186.344 | 11.247 | –21.418 |
| 232 | keptBest | –20.476 | 181.412 | 11.552 | –20.476 |
| 324 | Heredity | –20.359 | 185.809 | 11.279 | –20.359 |
| 79 | Heredity | –20.276 | 185.837 | 11.277 | –20.276 |
| 78 | softmutate | –20.162 | 180.825 | 11.59 | –20.162 |
| 196 | Random | –19.428 | 181.706 | 11.534 | –19.428 |
| 73 | keptBest | –19.278 | 180.853 | 11.588 | –19.278 |
| 123 | Permutate | –18.845 | 189.161 | 11.079 | –18.845 |
| 284 | Heredity | –18.6 | 188.341 | 11.127 | –18.6 |
| 246 | keptBest | –18.43 | 180.853 | 11.588 | –18.43 |
| 283 | Random | –18.119 | 182.093 | 11.509 | –18.119 |



# Additional Le Bail Refinements

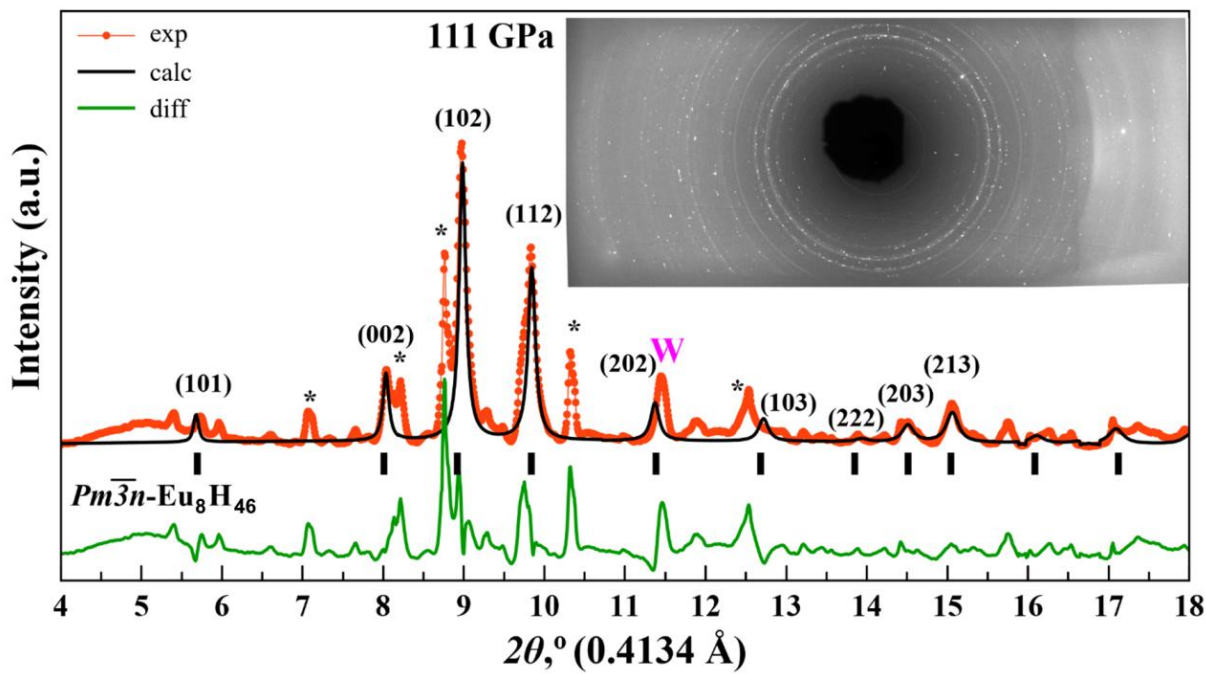

**Figure S12.** Experimental XRD pattern (cell #E1) and the Le Bail refinement of the main phase — $Pm\bar{3}n$-$Eu_8H_{46}$ at 111 GPa. Unexplained peaks are marked by asterisks.

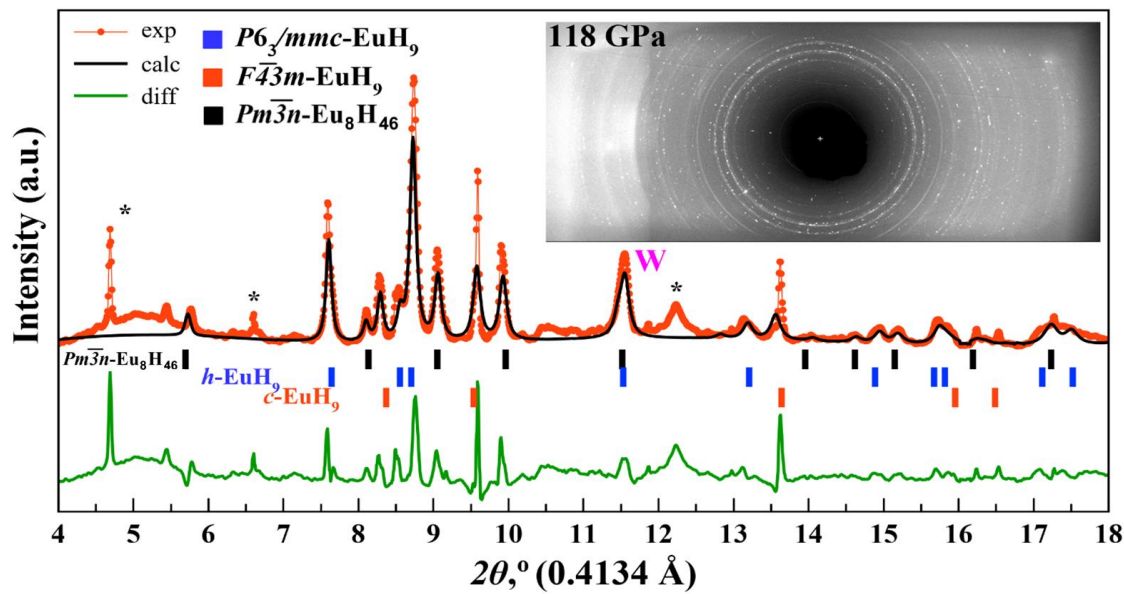

**Figure S13.** Experimental XRD pattern (cell #E1) and the Le Bail refinement of $F\bar{4}3m$-$EuH_9$, $P6_3/mmc$-$EuH_9$, and $Pm\bar{3}n$-$Eu_8H_{46}$ at 118 GPa. Unexplained peaks are marked by asterisks.



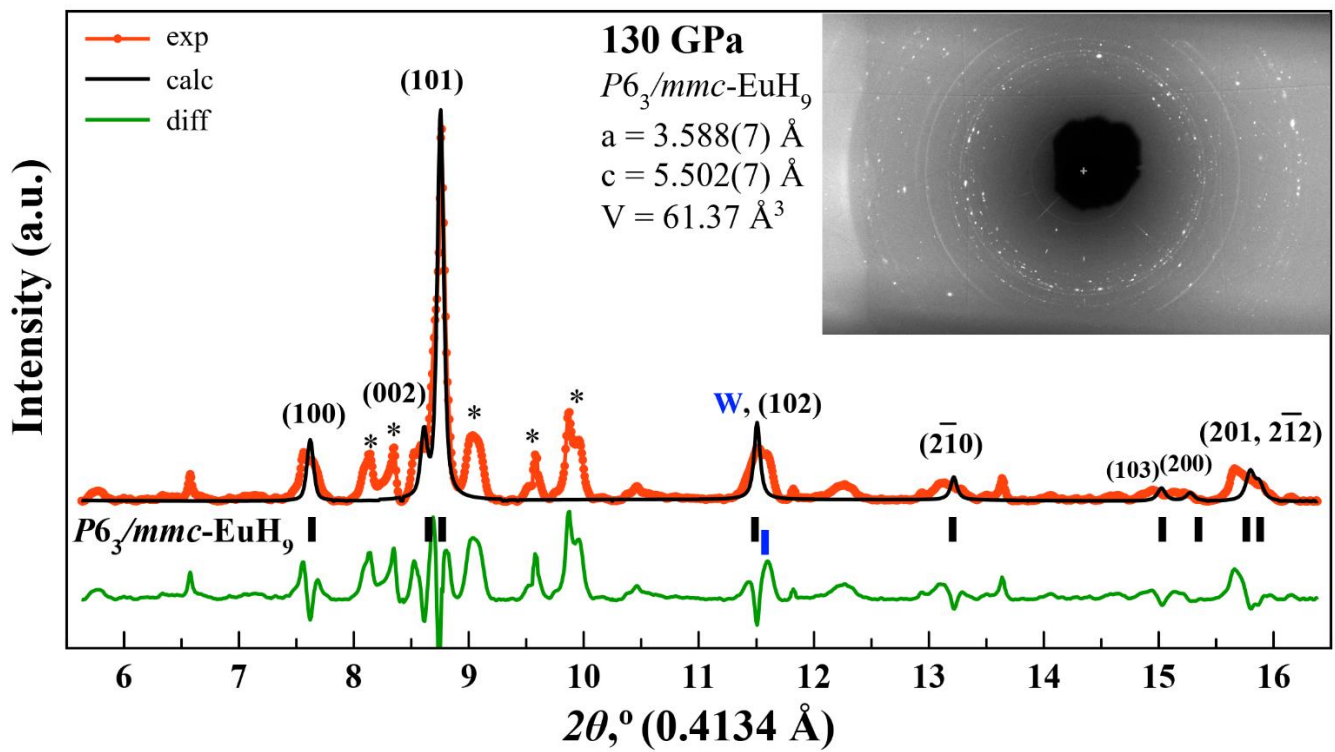

**Figure S14.** Experimental XRD pattern (cell #E1) and the Le Bail refinement of the main phase — $P6_3/mmc$-EuH$_9$ at 130 GPa. Unexplained peaks are marked by asterisks.

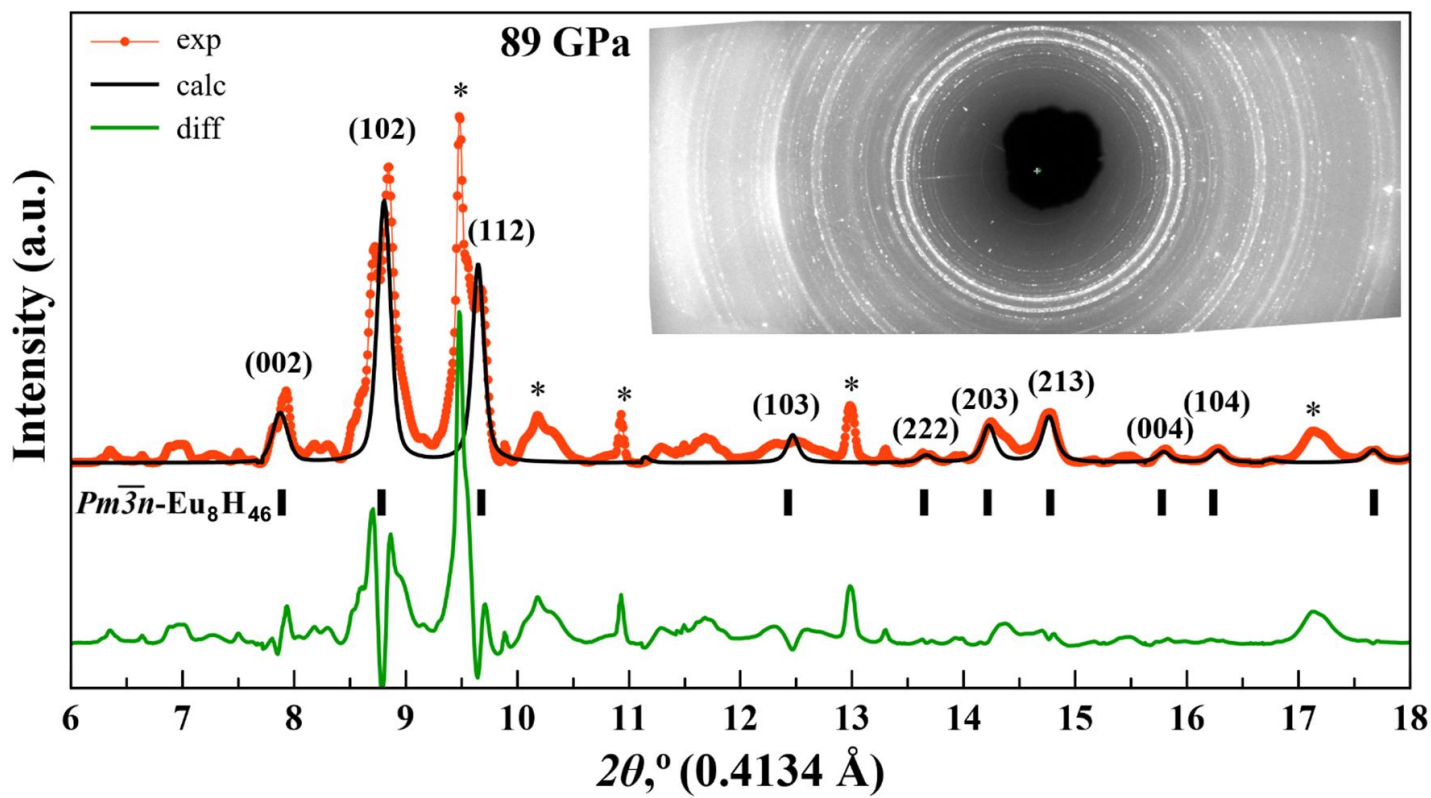

**Figure S15.** Experimental XRD pattern (cell #E2) and the Le Bail refinement of the main phase — $Pm\bar{3}n$-Eu$_8$H$_{46}$ at 89 GPa. Unexplained peaks are marked by asterisks.



# Magnetic Structure

The first step in a detailed study of the magnetic configurations of all europium hydrides is to determine which collinear magnetic ordering is preferable: ferromagnetic (FM) or antiferromagnetic (AFM). If we neglect the spin–orbit coupling, then only the magnitude of magnetic moments is meaningful and we should consider only one possible FM configuration, where the magnetic moments of the Eu atoms point in the same direction. The number of possible AFM configurations, instead, is in principle infinite, since it is always possible to find a new configuration, with half of the spins pointing up and the other half pointing down, by enlarging the unit cell. For this reason, we studied all possible AFM configurations only up to the $Eu_4H_{36}$ supercell for the $EuH_9$ stoichiometry. We generated our trial configurations using the derivative structure enumeration library *enumlib*,[39] in which the minimum size able to represent an AFM configuration is always chosen as the size of the unit cell. In other words, if we consider Eu atoms with different spin states as different atomic types, we turn out to always work with primitive cells. We obtained a total of seven AFM configurations for $F\bar{4}3m$-$EuH_9$ and five configurations for $P6_3/mmc$-$EuH_9$, as shown in Figure 4g,h.

Then we performed a full relaxation of both $EuH_9$ phases at 130 GPa without SOC for our eight (seven AFM and one FM) or six (five AFM and one FM) magnetic configurations. All of them successfully completed the process, but after relaxation some of the AFM configurations displayed distorted geometries, with symmetry and enthalpy lower than those of the ideal cubic or hexagonal phase (Table S9).

To explore the magnetic properties of $Pm\bar{3}n$-$Eu_8H_{46}$, we examined 15 trial configurations: one FM and 14 AFM (Figure S16). All of them completed relaxation without a change in the orientation of the magnetic moments. As in the case of $EuH_9$, some of the AFM configurations displayed a distorted geometry after relaxation (Table S11). We investigated the space group of all relaxed structures with increasing tolerances and found all distortions becoming negligible if a tolerance of 0.2 is used.

Magnetic anisotropy has been investigated by performing single-point energy calculations with SOC on a FM and an AFM configuration for each of our phases (ideal and distorted ones), with the magnetic moments aligned along seven different directions (Table S12). We expect a little magnetic anisotropy for all phases with $EuH_9$ stoichiometry, for which we identified the most stable orientation of the magnetic moments. For $Pm\bar{3}n$-$Eu_8H_{46}$ we do not expect any magnetic anisotropy since different orientations of the magnetic moments have almost the same enthalpy up to numerical errors.



**Table S11**. Enthalpy and space groups of the trial magnetic structures after full relaxation without SOC at 130 GPa. The space groups were determined with a tolerance of 0.1.

|  | Configuration | Enthalpy (meV/Eu atom) | Space group |
|---|---|---|---|
| $F\bar{4}3m$-EuH$_9$ | FM | 470.32 | $F\bar{4}3m$ |
|  | AFM1 | 0.0 | $P1$ |
|  | AFM2 | 440.52 | $F\bar{4}3m$ |
|  | AFM3 | 154.85 | $Imm2$ |
|  | AFM4 | 165.9 | $Imm2$ |
|  | AFM5 | 167.95 | $Imm2$ |
|  | AFM6 | 447.47 | $F\bar{4}3m$ |
|  | AFM7 | 180.47 | $Imm2$ |
| $P6_3/mmc$-EuH$_9$ | FM | 197.37 | $P6_3/mmc$ |
|  | AFM1 | 215.26 | $P6_3/mmc$ |
|  | AFM2 | 213.95 | $P6_3/mmc$ |
|  | AFM3 | 0.0 | $Cmcm$ |
|  | AFM4 | 59.26 | $C2$ |
|  | AFM5 | 8.2 | $Cmcm$ |
| $Pm\bar{3}n$-Eu$_8$H$_{46}$ | FM | 0.0 | $Pm\bar{3}n$ |
|  | AFM1 | 56.92 | $Pm\bar{3}n$ |
|  | AFM2 | 64.78 | $Pm\bar{3}n$ |
|  | AFM3 | 64.73 | $Pm\bar{3}n$ |
|  | AFM4 | 56.39 | $Pm\bar{3}n$ |
|  | AFM5 | 56.68 | $P4_2/mmc$ |
|  | AFM6 | 66.69 | $Pm\bar{3}n$ |
|  | AFM7 | 66.66 | $Pm\bar{3}n$ |
|  | AFM8 | 56.11 | $P4_2/mmc$ |
|  | AFM9 | 56.39 | $Pm\bar{3}n$ |
|  | AFM10 | 66.59 | $Pm\bar{3}n$ |
|  | AFM11 | 66.63 | $Pm\bar{3}n$ |
|  | AFM12 | 56.92 | $Pm\bar{3}n$ |
|  | AFM13 | 64.74 | $Pm\bar{3}n$ |
|  | AFM14 | 64.75 | $Pm\bar{3}n$ |



**Table S12.** Single-point enthalpies of FM and AFM configurations for both EuH$_9$ and $Pm\bar{3}n$-Eu$_8$H$_{46}$ (including the distorted structures $P$1, $Imm$2, $Cmcm$, $C$2), calculated with SOC and with the magnetic moments aligned along seven different directions at 130 GPa.

| Configuration | Enthalpy (meV/atom) | | | | | | |
|---|---|---|---|---|---|---|---|
| $P$1(cub)-EuH$_9$ | X | Y | Z | XY | XZ | YZ | XYZ |
| FM | - | 1.047* | - | - | 0.000* | 0.532* | 0.791* |
| AFM1 | - | 0.617 | - | 0.158* | - | 0.430 | 0.115 |
| $Imm$2(cub)-EuH$_9$ | | | | | | | |
| FM | 26.928 | 26.651 | 27.085 | 26.78 | 26.994 | 26.277 | 26.471 |
| AFM3 | 0.871 | 0.833 | 1.392 | 0.849 | 1.131 | **0.0** | 0.291 |
| $F\bar{4}3m$-EuH$_9$ | | | | | | | |
| FM | 30.226 | 30.283 | 30.275 | 30.272 | 30.244 | 30.318 | 30.281 |
| AFM2 | **0.0** | 0.41 | 0.205 | 0.171 | 0.076 | 0.652 | 0.397 |
| $Cmcm$ (hex)-EuH$_9$ | | | | | | | |
| FM | 7.691 | 8.878 | 7.309 | 8.289 | 7.489 | 8.12 | 7.958 |
| AFM3 | 0.076 | 1.4 | **0.0** | 0.722 | 0.01 | 0.827 | 0.548 |
| $C$2(hex)-EuH$_9$ | | | | | | | |
| FM | 0.563 | **0.0** | 0.396 | 0.296 | 0.767 | 0.185 | 0.498 |
| AFM4 | 6.517 | 5.65 | 6.018 | 6.091 | 6.386 | 5.816 | 6.12 |
| $P6_3/mmc$-EuH$_9$ | | | | | | | |
| FM | 0.176 | 0.176 | **0.0** | 0.174 | 0.118 | 0.128 | 0.152 |
| AFM1 | 19.372 | 19.371 | 18.4 | 19.368 | 18.855 | 18.874 | 19.028 |
| $Pm\bar{3}n$-Eu$_8$H$_{46}$ | | | | | | | |
| FM | 0.057 | 0.057 | 0.057 | 0.023 | 0.022 | 0.016 | **0.0** |
| AFM | 57.988 | 58.035 | 58.035 | 58.009 | 58.009 | 58.021 | 57.999 |

"-" means the convergence was not reached after 200 electronic steps.

* Magnetic moments changed the orientation.



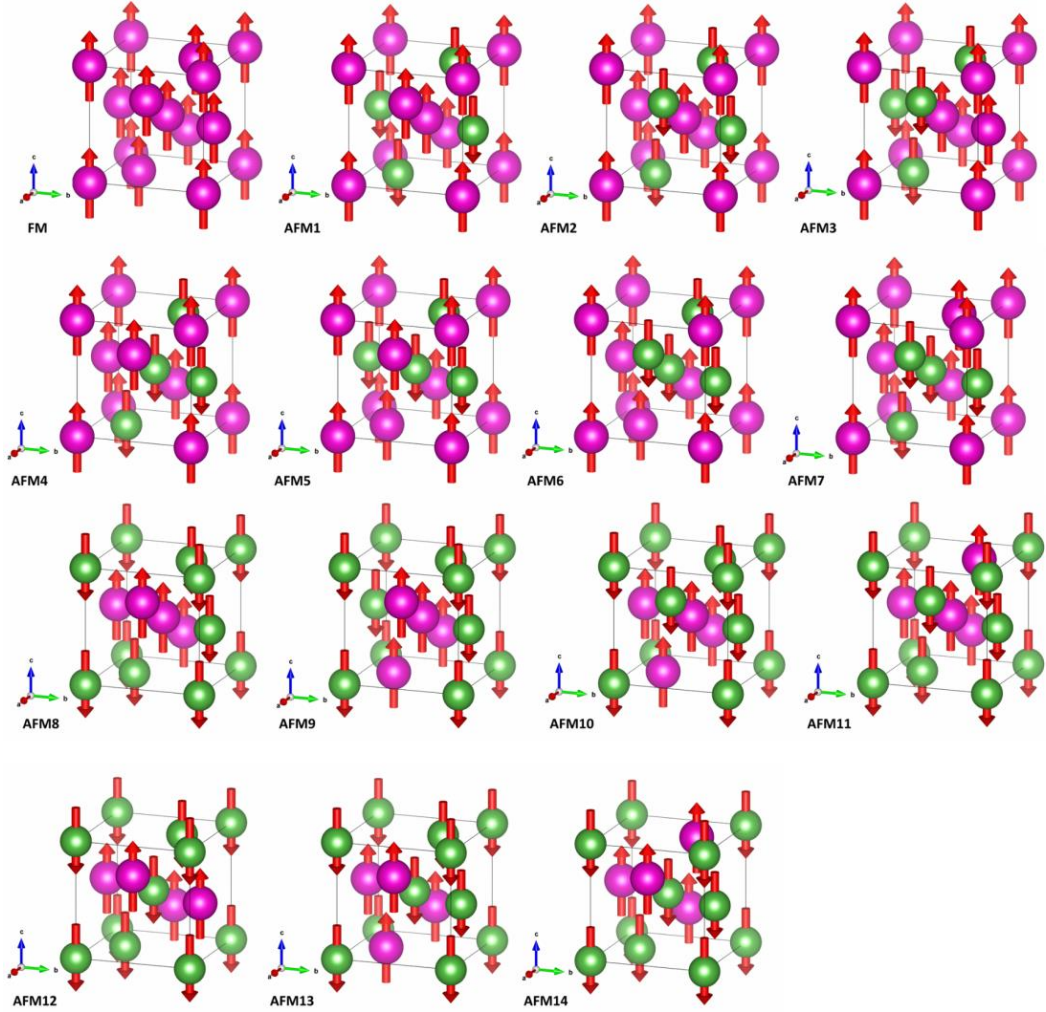

**Figure S16.** 15 trial magnetic configurations for $Pm\bar{3}n$-Eu$_8$H$_{46}$ used in this work.

To study the Néel temperature of $F\bar{4}3m$-EuH$_9$ and the Curie temperature of $P6_3/mmc$-EuH$_9$, we relaxed two additional independent AFM configurations for each geometry with SOC and the magnetic moments aligned along the $x$ axis and $z$ axis, respectively. We used magnetic moments aligned along the $z$ axis to study the Curie temperature of ferromagnetic $Pm\bar{3}n$-Eu$_8$H$_{46}$. The final enthalpies and magnetic moment orientations are shown in Table S13, whereas the absolute value of the magnetic moment on each Eu atom after relaxation was always close to 6.8 $\mu_B$ for EuH$_9$ and 6.9 $\mu_B$ for Eu$_8$H$_{46}$.

Taking into account the crystal periodicity, in the unit cell of $F\bar{4}3m$-EuH$_9$ we found 24 links between the Eu atoms at 3.51 Å and 12 links at 4.96 Å at 130 GPa. In the unit cell of $P6_3/mmc$-EuH$_9$, we found 12 links at 3.49 Å and 12 links at 3.55 Å. Then we modeled the magnetic interaction with the Ising Hamiltonian:

$$H = H_0 - \frac{1}{2}J_1 \sum_{i \neq j} S_i \cdot S_j - \frac{1}{2}J_2 \sum_{i \neq j} S_i \cdot S_j \tag{S1}$$

where $J_1$ and $J_2$ are the coupling constants assigned to the two groups of neighbors, $H_0$ is a constant term which does not depend on the magnetic interaction, and $S_i$ is +1 or –1, depending on the spin orientation of the respective Eu atom. This Hamiltonian for EuH$_9$ takes into account the magnetic interactions only up to the second nearest neighbors.



**Table S13.** Magnetic moment orientations and calculated enthalpies of the three magnetic configurations used for studying the Néel temperature of EuH$_9$ and of the four magnetic configurations used for studying the Curie temperature of $Pm\bar{3}n$-Eu$_8$H$_{46}$. +/− mark the direction of magnetic moments along the $z$ axis.

| | \multicolumn{5}{c}{$F\bar{4}3m$-EuH$_9$} |
|---|---|---|---|---|---|
| | Eu1 | Eu2 | Eu3 | Eu4 | Enthalpy (meV/Eu atom) |
| FM | + | + | + | + | 31.2 |
| AFM2 | + | + | − | − | 0.0 |
| AFM5 | + | − | + | − | 8.3 |
| | \multicolumn{5}{c}{$P6_3/mmc$-EuH$_9$} |
| | Eu1 | Eu2 | Eu3 | Eu4 | Enthalpy (meV/Eu atom) |
| FM | + | + | + | + | 0.0 |
| AFM1 | + | − | + | − | 20.0 |
| AFM3 | + | + | − | − | 16.7 |
| | \multicolumn{9}{c}{$Pm\bar{3}n$-Eu$_8$H$_{46}$} |
| | Eu1 | Eu2 | Eu3 | Eu4 | Eu5 | Eu6 | Eu7 | Eu8 | Enthalpy (meV/Eu atom) |
| FM | + | + | + | + | + | + | + | + | 0.0 |
| AFM1 | + | + | + | + | − | − | − | − | 58.0 |
| AFM5 | + | − | + | + | − | − | − | + | 57.0 |
| AFM8 | + | − | + | − | + | − | + | − | 66.9 |

In the primitive cell of Eu$_8$H$_{46}$, we found six links between the Eu atoms at 2.95 Å, 24 links at 3.29 Å, and 24 links at 3.61 Å. We modeled the magnetic interaction with the Ising Hamiltonian:

$$H = H_0 - \frac{1}{2}J_1 \sum_{i \neq j} S_i \cdot S_j - \frac{1}{2}J_2 \sum_{i \neq j} S_i \cdot S_j - \frac{1}{2}J_3 \sum_{i \neq j} S_i \cdot S_j \tag{S2}$$

where $J_1$, $J_2$, and $J_3$ are the coupling constants assigned to the three groups of neighbors mentioned above. This Ising Hamiltonian takes into account the magnetic interactions up to the 3rd nearest neighbors for Eu$_8$H$_{46}$.

The values of the coupling constants of each phase can be obtained from the enthalpies of our three magnetic configurations by solving the system of equations for $H_0$, $J_1$, $J_2$, and $J_3$. The critical temperatures in our Heisenberg model were obtained from a Monte Carlo simulation as implemented in the VAMPIRE code.[40] The size of the simulation box was 8×8×8 nm for EuH$_9$ and 10×10×10 nm for Eu$_8$H$_{46}$ after the convergence tests, and the transition temperature was found as the value where the normalized mean magnetization length goes below 0.25. The obtained values of the coupling constants and the Néel or Curie temperatures are listed in Tables S14-16, the mean magnetization length is plotted vs. temperature in Figure S17. Obtained normalized μ(T) dependencies may be approximated well by the Curie–Weiss type formula $a+b \cdot (T_{C,N} - T)^{1/2}$, with b = 0.115 ($c$-EuH$_9$), 0.066 ($h$-EuH$_9$) and 0.042 (Eu$_8$H$_{46}$).

**Table S14.** Coupling constants (in Joules) and the estimated critical temperatures of both EuH$_9$ phases and $Pm\bar{3}n$-Eu$_8$H$_{46}$.

| | $J_1$ (J per Eu–Eu link) | $J_2$ (J per Eu–Eu link) | $J_3$ (J per Eu–Eu link) | $T_{C,N}$ (K) |
|---|---|---|---|---|
| $F\bar{4}3m$-EuH$_9$ | −6.2578e−22 | 1.8056e−23 | - | 24 |
| $P6_3/mmc$-EuH$_9$ | 2.6834e−22 | 5.3336e−22 | - | 137 |
| $Pm\bar{3}n$-Eu$_8$H$_{46}$ | 2.5042e−21 | 1.4070e−21 | 9.1531e−22 | 336 |



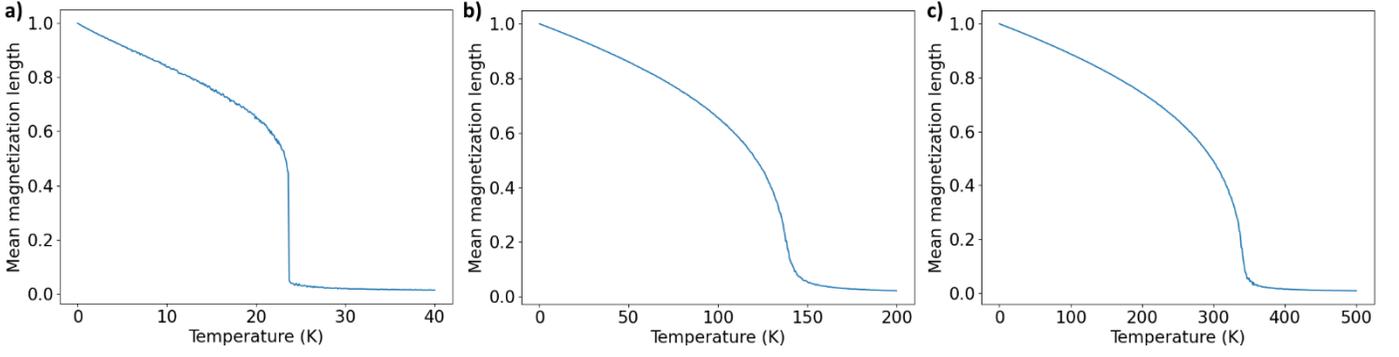

**Figure S17.** Normalized mean magnetization lengths with respect to temperature in (a) $F\bar{4}3m$-EuH$_9$, (b) $P6_3/mmc$-EuH$_9$, and (c) $Pm\bar{3}n$-Eu$_8$H$_{46}$ from the Monte Carlo simulations. For antiferromagnetic $F\bar{4}3m$-EuH$_9$ the mean magnetization length of the spin-up channel is displayed.

We chose to truncate the Ising Hamiltonian for the EuH$_9$ phases to 2nd nearest neighbor interactions on the basis of a convergence test. For $F\bar{4}3m$-EuH$_9$, we first calculated $J_1$ from all combinations of enthalpies listed in Table S13, obtaining quite similar values (Table S15). The extension of our Ising Hamiltonian to the 2nd neighbor interactions was dictated only by the possibility of increasing the precision. For $P6_3/mmc$-EuH$_9$, the values of $J_1$ for all possible combinations of calculated enthalpies are displayed in Table S15. In one case the system of equations was singular, and in the two remaining cases the calculated values of $J_1$ are very different. The negative value of $J_1$ is unphysical because it denotes an AFM ordering, but our system is ferromagnetic. Therefore, we chose to include the second neighbor interactions in our Ising Hamiltonian. We expected them to be of crucial importance in $P6_3/mmc$-EuH$_9$ because they are equally abundant and their distance $d$(Eu–Eu) is only 0.06 Å larger compared with the first neighbor interactions. We did not include the 3rd neighbor interactions because their distance, 4.98 Å, is significantly higher than that of the first and the second neighbor interactions and they are equally abundant in the unit cell.

**Table S15**. Values of $J_1$ (in Joules) for both EuH$_9$ phases and Eu$_8$H$_{46}$, obtained from different combinations of magnetic configurations.

| Phase | Combination | $J_1$ (J per Eu-Eu link) |
| --- | --- | --- |
| $F\bar{4}3m$-EuH$_9$ | FM, AFM2 | –6.2578e–22 |
| | FM, AFM5 | –6.1375e–22 |
| | AFM2, AFM5 | –6.6190e–22 |
| $P6_3/mmc$-EuH$_9$ | FM, AFM2 | singular |
| | FM, AFM5 | 6.6836e–22 |
| | AFM2, AFM5 | –1.3168e–22 |
| $Pm\bar{3}n$-Eu$_8$H$_{46}$ | FM, AFM1 | singular |
| | FM, AFM5 | 1.8269e–20 |
| | FM, AFM8 | 7.1488e–21 |
| | AFM1, AFM5 | –3.0969e–22 |
| | AFM1, AFM8 | 9.5606e–22 |
| | AFM5, AFM8 | 1.5889e–21 |

The choice to truncate the Ising Hamiltonian for Eu$_8$H$_{46}$ to the 3rd nearest neighbor interactions is also the result of a convergence test. In principle, we could have used only the nearest neighbors, calculating $J_1$ from any



combination of two enthalpies from Table S13 that gives a nonsingular system of equations. The resulting values of $J_1$ are presented in Table S15. The results are very dependent on the choice of the systems, and in one case we got a negative value of $J_1$, which is unphysical because it denotes an AFM ordering, whereas our system is ferromagnetic. In a similar fashion, we could have truncated our Ising Hamiltonian at the 2nd nearest neighbors. In this case we need three relaxed magnetic configurations to calculate $J_1$ and $J_2$. The results for all possible choices are listed in Table S16 together with the respective Curie temperatures from Monte Carlo simulations.

**Table S16**. Values of $J_1$, $J_2$ (in Joules) and $T_C$ for $Eu_8H_{46}$ obtained from different combinations of magnetic configurations.

| Phase | Combination | $J_1$ (J per Eu–Eu link) | $J_2$ (J per Eu–Eu link) | $T_C$ (K) |
|---|---|---|---|---|
| $Pm\bar{3}n$-$Eu_8H_{46}$ | FM, AFM1, AFM5 | 4.3349e–21 | 2.3323e–21 | 376 |
| | FM, AFM1, AFM8 | 2.5042e–21 | 2.3223e–21 | 338 |
| | FM, AFM5, AFM8 | 1.5889e–21 | 2.7799e–21 | 374 |
| | AFM1, AFM5, AFM8 | 1.5889e–21 | 9.4931e–22 | 152 |

For the first three combinations, the obtained values of $J_2$ are very close to each other. The 2nd nearest neighbor interaction in $Pm\bar{3}n$-$Eu_8H_{46}$ turns out to be more important than the nearest neighbor interaction because the values of the coupling constants $J_1$ and $J_2$ are comparable and the number of second nearest neighbor interactions is four times higher. This is the main reason why the first three Curie temperatures are also close. The results in the last row of Table S14 are most probably due to numerical errors because the enthalpies of the three AFM configurations are relatively close to each other (Table S11). For our calculation of $T_C$ for $Eu_8H_{46}$, we chose to take another step forward in accuracy by writing our Ising Hamiltonian up to the 3rd nearest neighbors.

A somewhat more complicated case is pseudohexagonal $Cmcm$-$EuH_9$ (Table S12). In order to study the Néel temperature of this phase we checked firstly how much the distances between neighbouring atoms changed with respect to the high-symmetry $P6_3/mmc$ modification. For computing the Curie temperature of FM $P6_3/mmc$-$EuH_9$, we wrote the Ising Hamiltonian up to the second nearest neighbors. The situation is different for $Cmcm$-$EuH_9$, where the first and the second neighbours' distances split into three groups. In the $Cmcm$-$Eu_4H_{36}$ supercell there are 8 Eu-Eu links at 3.47 Å, 8 links at a distance between 3.52 Å and 3.55 Å and 8 links at 3.61 Å. The next group of neighbours are located at more than 5 Å and therefore we did not take them into account. We figured out that this structural feature of the $Cmcm$ is responsible for the increased stability of the AFM3 (Figure 4h) magnetic state compared to the FM state of $P6_3/mmc$ modification.

In the $Cmcm$-$Eu_4H_{36}$ two more AFM configurations are possible other than AFM3. Together with the FM state, four magnetic configurations are enough to compute three coupling constants ($J_1$, $J_2$, $J_3$) only if they give rise to a non-singular system of equations. Unfortunately, this is not the case for $Cmcm$ phase, therefore the coupling constants were computed by using the FM state and three independent AFM configurations in the bigger supercell $Eu_8H_{72}$. As a result we found: $J_1 = 4.7416 \times 10^{-22}$ J, $J_2 = 1.1398 \times 10^{-21}$ J, $J_3 = -1.9388 \times 10^{-21}$ J.

Monte Carlo simulation on the AFM3 magnetic state in the unit cell $Cmcm$-$Eu_4H_{36}$ using found coupling constants indicates that the AFM3 magnetic ordering is stable only at 0 K, while it collapses at any finite temperature. Among all 76 possible AFM configurations in the supercell $Eu_8H_{72}$ we found that one configuration, denoted here as AFM10, is more stable then the previously described AFM3. However, the Monte Carlo simulation on AFM10 gives the same result as for AFM3. Thus, we can conclude that $Cmcm$-$EuH_9$ does not exhibit magnetism at finite temperatures. This conclusion is very important from the point of view of a possible manifestation of superconductivity in pseudohexagonal $EuH_9$.



# Natural Population Analysis

Table S17. Occupancies (|e|) of localized bonding elements obtained via the SSAdNDP method.

|  | $P6_3/mmc$-CeH$_9$ (100 GPa) | $F\bar{4}3m$-PrH$_9$ (100 GPa) | $P6_3/mmc$-PrH$_9$ (100 GPa) | $P6_3/mmc$-NdH$_9$ (100 GPa) | $Fm\bar{3}m$-LaH$_{10}$ (150 GPa) | $Fm\bar{3}m$-YH$_{10}$ (400 GPa) |
|---|---|---|---|---|---|---|
| 1c–2e (s-type lone pairs) | 1.84 | 1.84 | 1.84 | 1.85 | 1.71 | 1.96 |
| Three 1c–2e (p-type lone pairs) | 1.94 | 1.95 | 1.95 | 1.95 | 1.89 | 1.98 |
| Five $n$c–2e* | 1.35–1.25 | 1.44–1.16 | 1.38–1.29 | 1.44–1.21 | 1.65–1.54 | 1.66–1.56 |
| One $n$c–2e* | 1.17 | 1.08 | 1.23 | 1.19 | 1.48 | 1.51 |

* $n$ is the number of the hydrogen atoms in the cage.

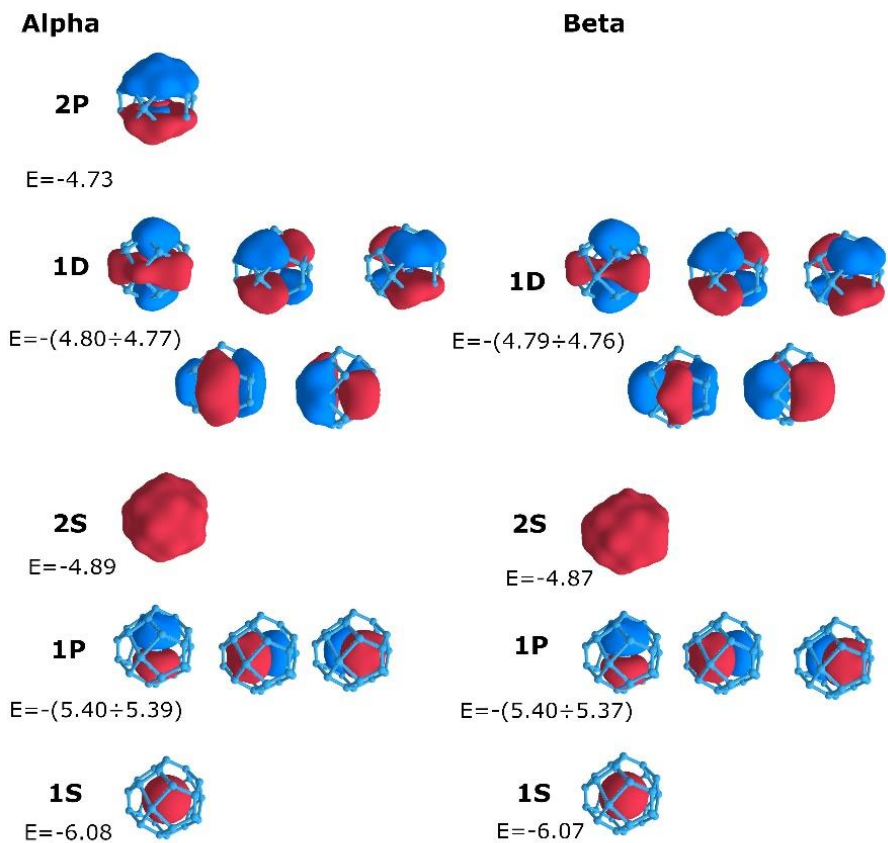

Figure S18. Molecular orbitals and energies (Hartree) of CeH$_{29}^{20+}$ cluster at 100 GPa.

Table S18. Average Bader charges in the investigated europium polyhydrides at 130 GPa.

|  | $Pm\bar{3}n$-Eu$_8$H$_{46}$ | $F\bar{4}3m$-EuH$_9$ | $P6_3/mmc$-EuH$_9$ |
|---|---|---|---|
| Average charge on Eu atoms | +1.066 | +1.086 | +1.147 |
| Average charge on H atoms | –0.185 | –0.121 | –0.127 |



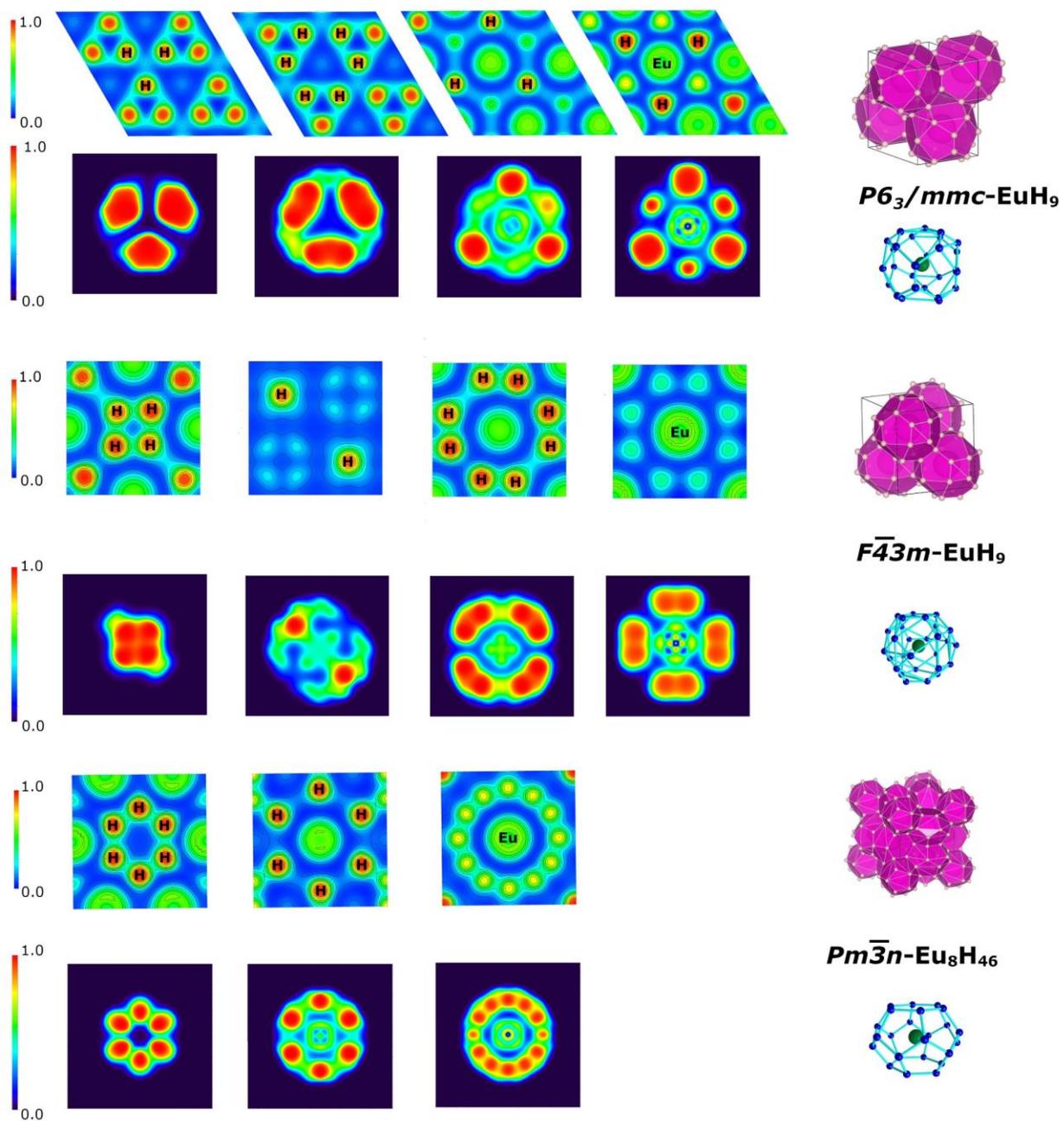

**Figure S19.** Comparison of the ELF plots of the bulk structures of EuH$_9$ and Eu$_8$H$_{46}$ and model clusters at 130 GPa.



# Electronic Band Structure

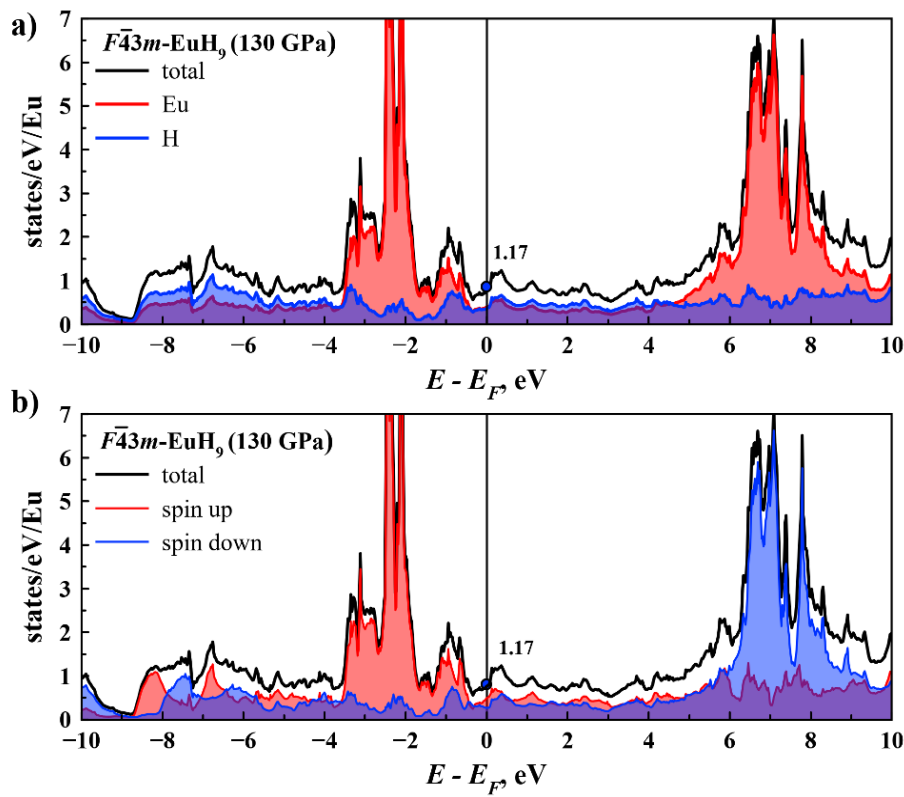

**Figure S20**. Electron density of states (DOS) of $F\bar{4}3m$-EuH$_9$ at 130 GPa per 1 Eu atom (DFT+U). (a) Contributions of the Eu and H atoms to the total DOS. (b) Contributions of different spin orientations to the total DOS.

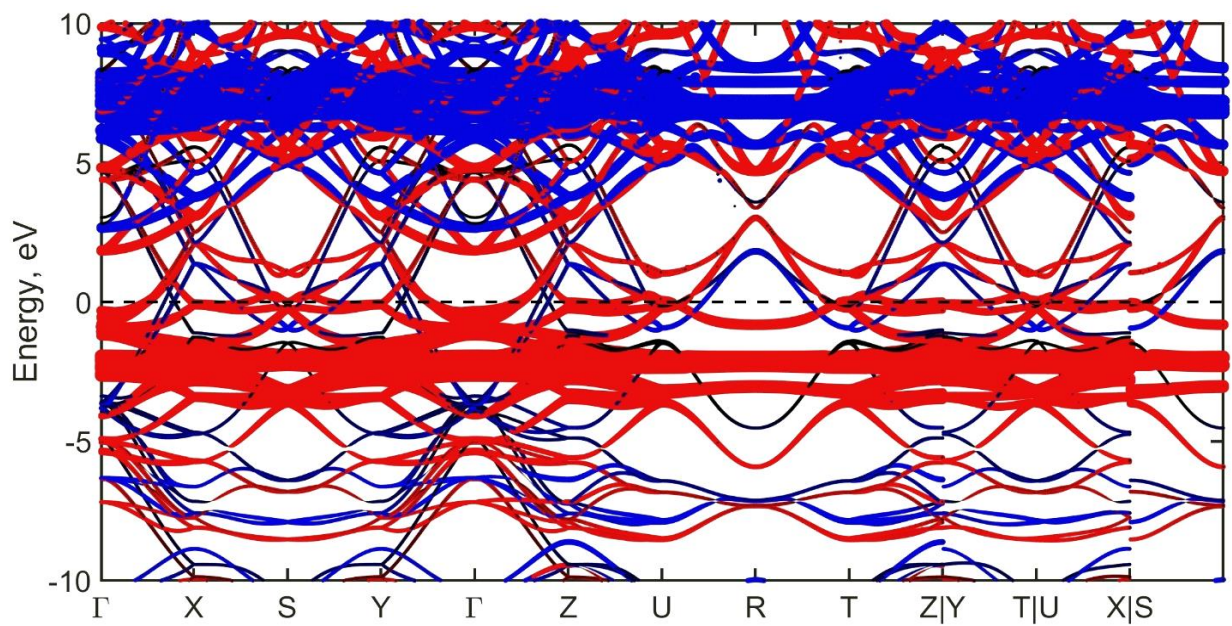

**Figure S21**. Spin-resolved electron band structure (DFT+U) of $F\bar{4}3m$-EuH$_9$ at 130 GPa. Bands of Eu are shown in red (spin up) and blue (spin down), bands of H - in black.



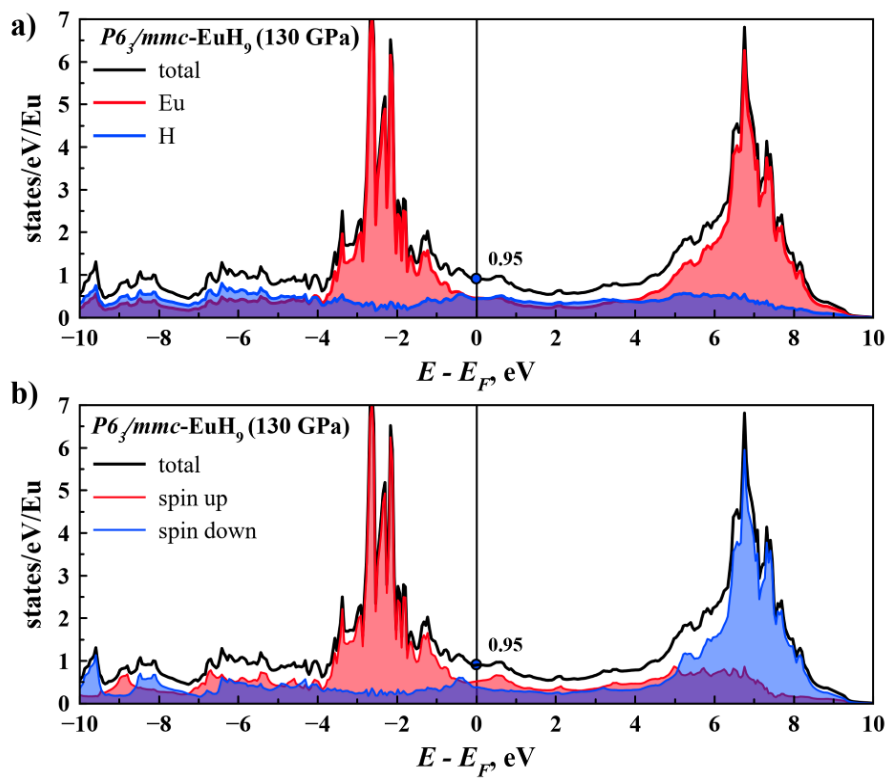

**Figure S22.** Electron density of states (DOS) of $P6_3/mmc$-EuH$_9$ at 130 GPa per 1 Eu atom. (a) Contributions of the Eu and H atoms to the total DOS. (b) Contributions of different spin orientations to the total DOS.

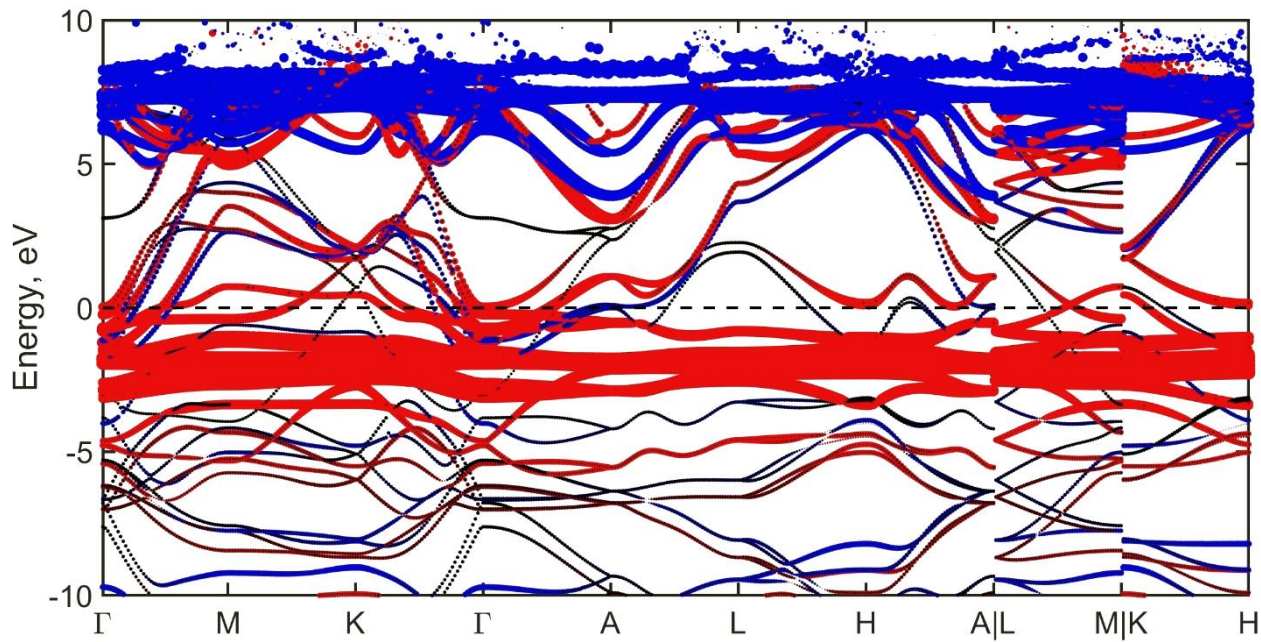

**Figure S23.** Spin-resolved electron band structure of $P6_3/mmc$-EuH$_9$ at 130 GPa. Bands of Eu are shown in red (spin up) and blue (spin down), bands of H presents in black.



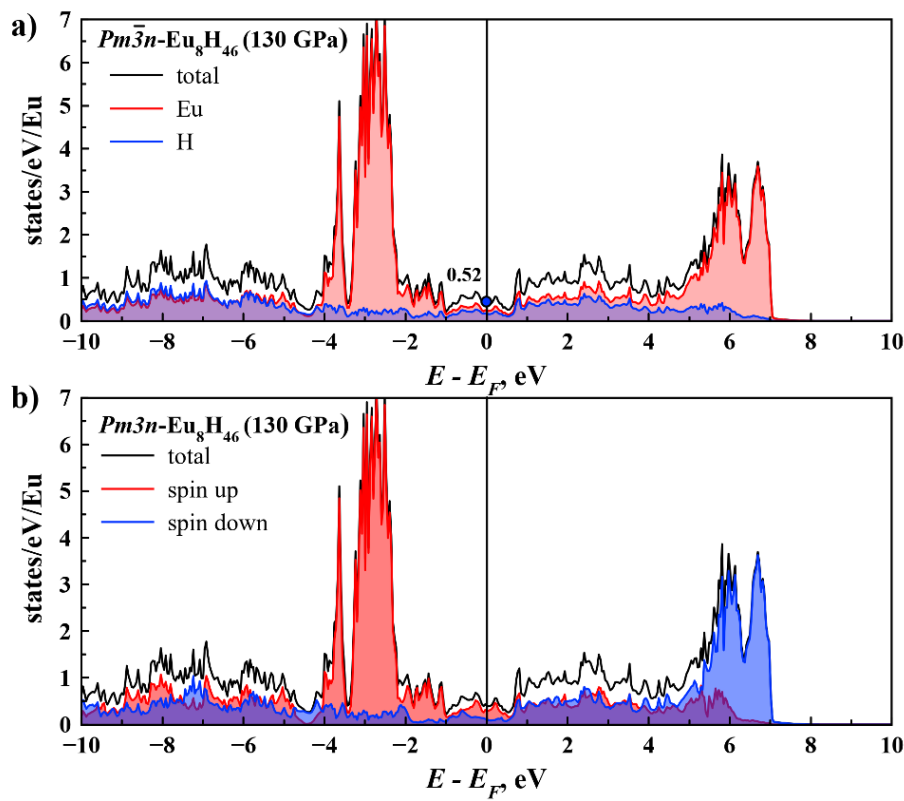

**Figure S24.** Electron density of states (DOS) of $Pm\bar{3}n$-$Eu_8H_{46}$ at 130 GPa per 1 Eu atom. (a) Contributions of the Eu and H atoms to the total DOS. (b) Contributions of different spin orientations to the total DOS.

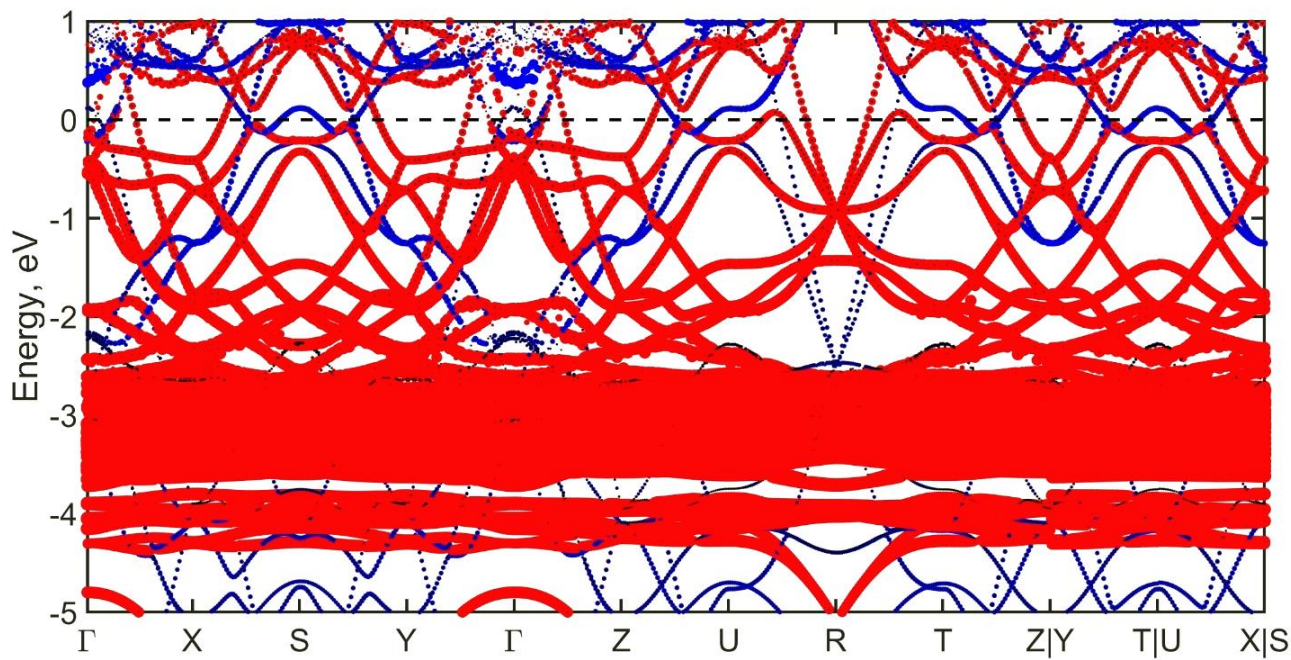

**Figure S25.** Spin-resolved electron band structure of $Pm\bar{3}n$-$Eu_8H_{46}$ at 130 GPa. Bands of Eu are shown in red (spin up) and blue (spin down), bands of H presents in black.



# Phonon Spectra

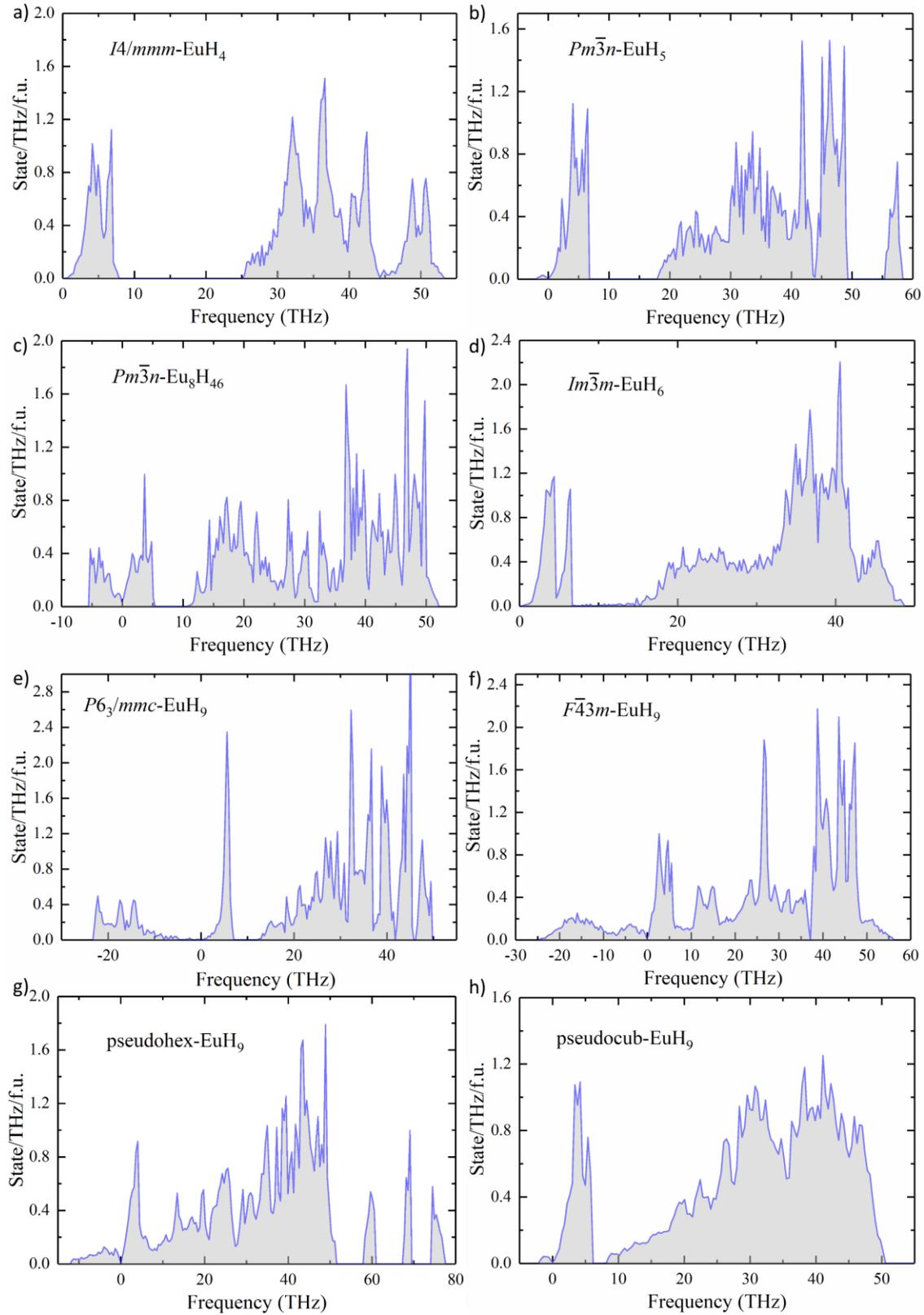

**Figure S26.** Phonon density of states for various Eu–H phases at 130 GPa with SOC and corresponding $U$–$J$ (Table S7). Almost all spectra have an imaginary part associated with the distortion of ideal high-symmetry structures or with bad convergence within the selected pseudopotentials. Ideal EuH$_9$ phases are unstable, whereas distorted $P1$-EuH$_9$ has much smaller number of imaginary frequencies. Ideal $Pm\bar{3}n$-Eu$_8$H$_{46}$ also undergoes distortion.


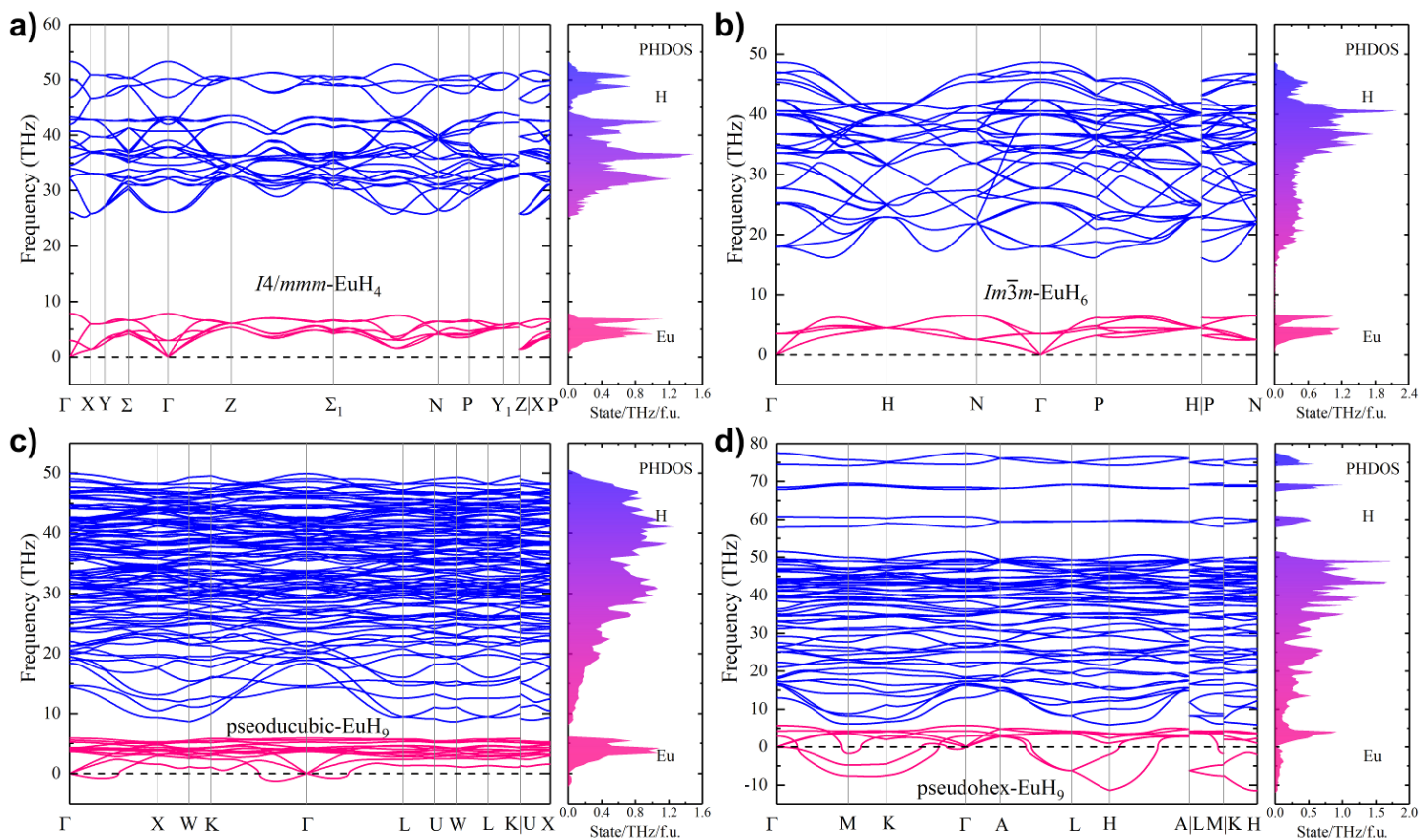

**Figure S27.** Phonon band structure of various Eu hydrides at 130 GPa with SOC and corresponding $U$–$J$ (Table S7). Almost all spectra have an imaginary part associated with distortion of ideal high-symmetry structures or with bad convergence within the selected PPs.

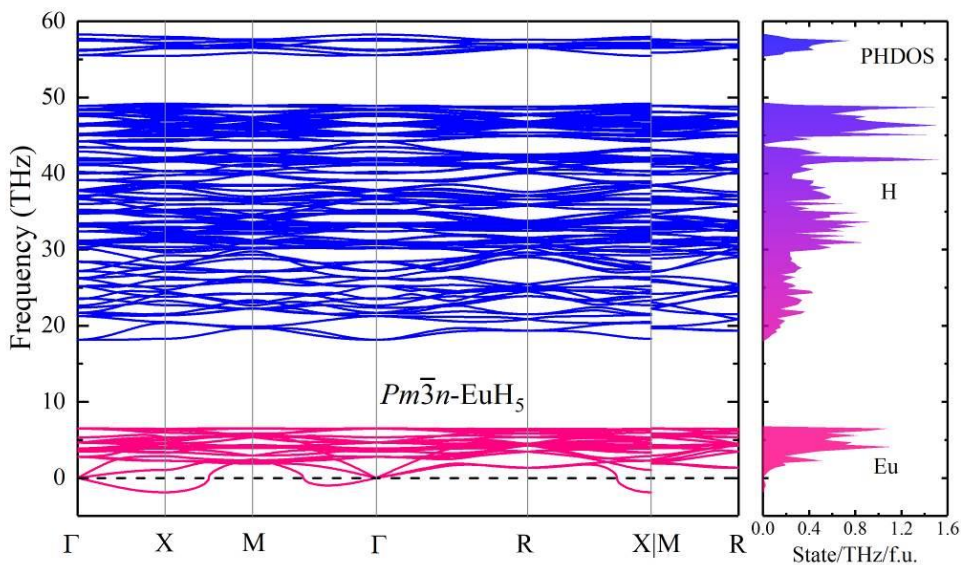

**Figure S28.** Phonon band structure of $Pm\bar{3}n$-EuH$_5$ at 130 GPa with SOC and corresponding $U$–$J$ (Table S7).



## Anharmonic Calculations

The dynamic stability and phonon density of states of the europium hydrides were studied using classical molecular dynamics and interatomic potential based on machine learning. We used the Moment Tensor Potential (MTP).[41] It has been demonstrated that the MTP can be used to calculate the phonon properties of materials.[41] Moreover, within this approach we can explicitly take into account the anharmonicity of hydrogen vibrations.

To train the potential, we first simulated studied europium hydrides in quantum molecular dynamics in an NPT ensemble at 130 GPa and 2000 K, with a duration of 5 picoseconds using the VASP code.[13–15] We used the PAW PBE pseudopotentials for the H and Eu atoms, $2\pi \times 0.06$ Å$^{-1}$ $k$-mesh with a cutoff energy of 400 eV, and a $2 \times 2 \times 1$ supercell with 80 atoms.

For training of the MTP, sets of both EuH$_9$ and Eu$_8$H$_{46}$ structures were chosen using active learning.[42] We checked the dynamical stability of studied europium hydrides with the obtained MTPs via several runs of molecular dynamics calculations at 300 K and 130 GPa. First, the NPT dynamics simulations were performed in a supercell with 960 atoms for 40 picoseconds. During the last 20 picoseconds, the cell parameters were averaged. At the second step, the coordinates of the atoms were averaged within the NVT dynamics with a duration of 20 picoseconds and the final structure was symmetrized. The space groups of both EuH$_9$ and Eu$_8$H$_{46}$ were retained, only the lattice parameters were slightly changed.

Then, for the structures of europium polyhydrides relaxed at 130 GPa and 300 K, the phonon density of states was calculated within the MTP using the velocity autocorrelator (VACF) separately for each type of atoms:[43]

$$g(\vartheta) = 4 \int_0^\infty \cos(2\pi\vartheta t) \frac{\langle \overline{\vartheta(0)\vartheta(t)} \rangle}{\langle \overline{\vartheta(0)^2} \rangle} dt \qquad (S3)$$

where $\vartheta$ is the frequency. The calculations were carried out in a $20 \times 20 \times 20$ supercell. The velocity autocorrelator was calculated using molecular dynamics, then the phonon DOS was obtained (Figure S29-31).

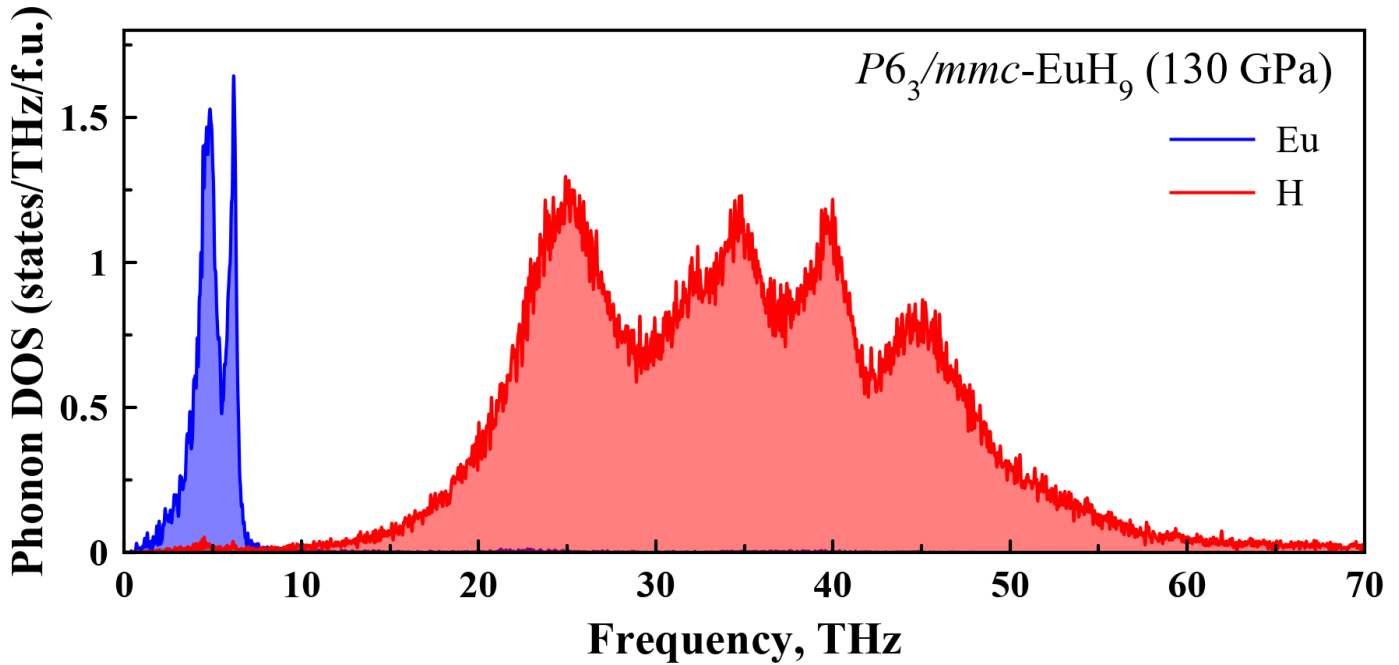

**Figure S29.** Anharmonic phonon density of states of $P6_3/mmc$-EuH$_9$ at 130 GPa and 300 K.



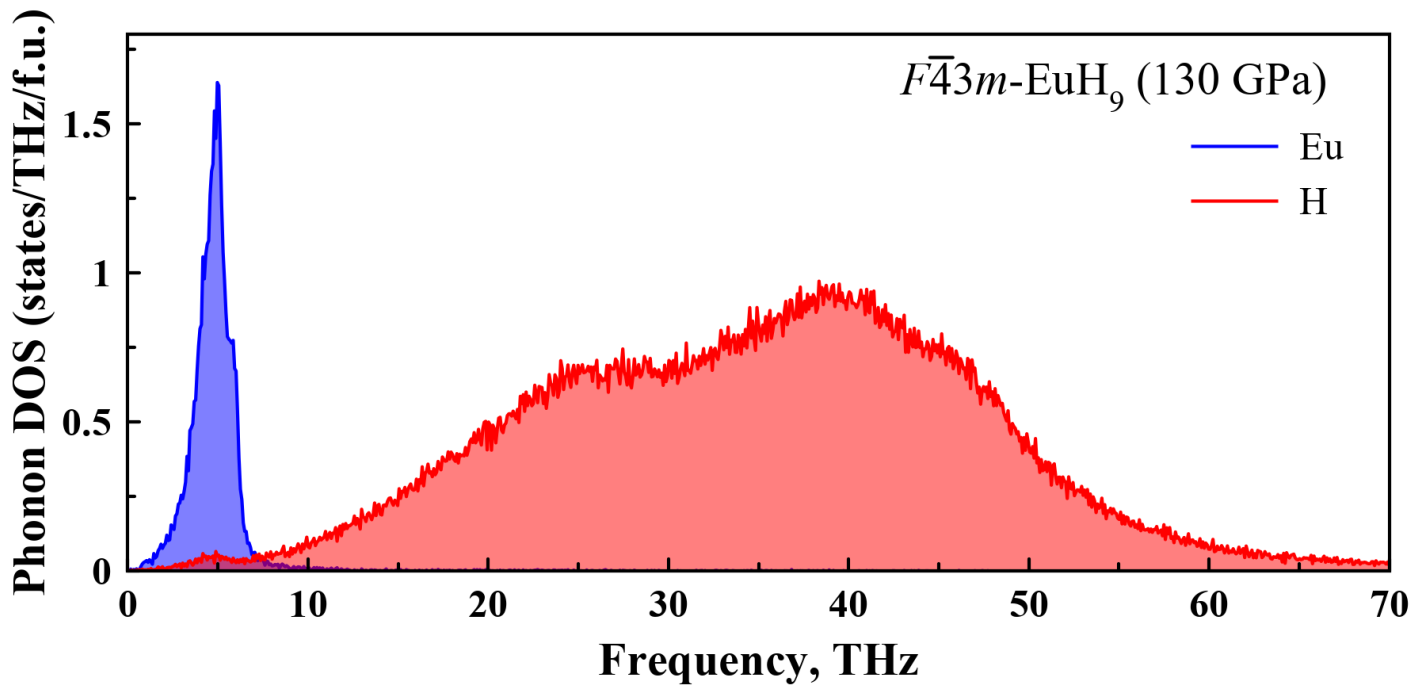

**Figure S30.** Anharmonic phonon density of states of $F\bar{4}3m$-EuH$_9$ at 130 GPa and 300 K.

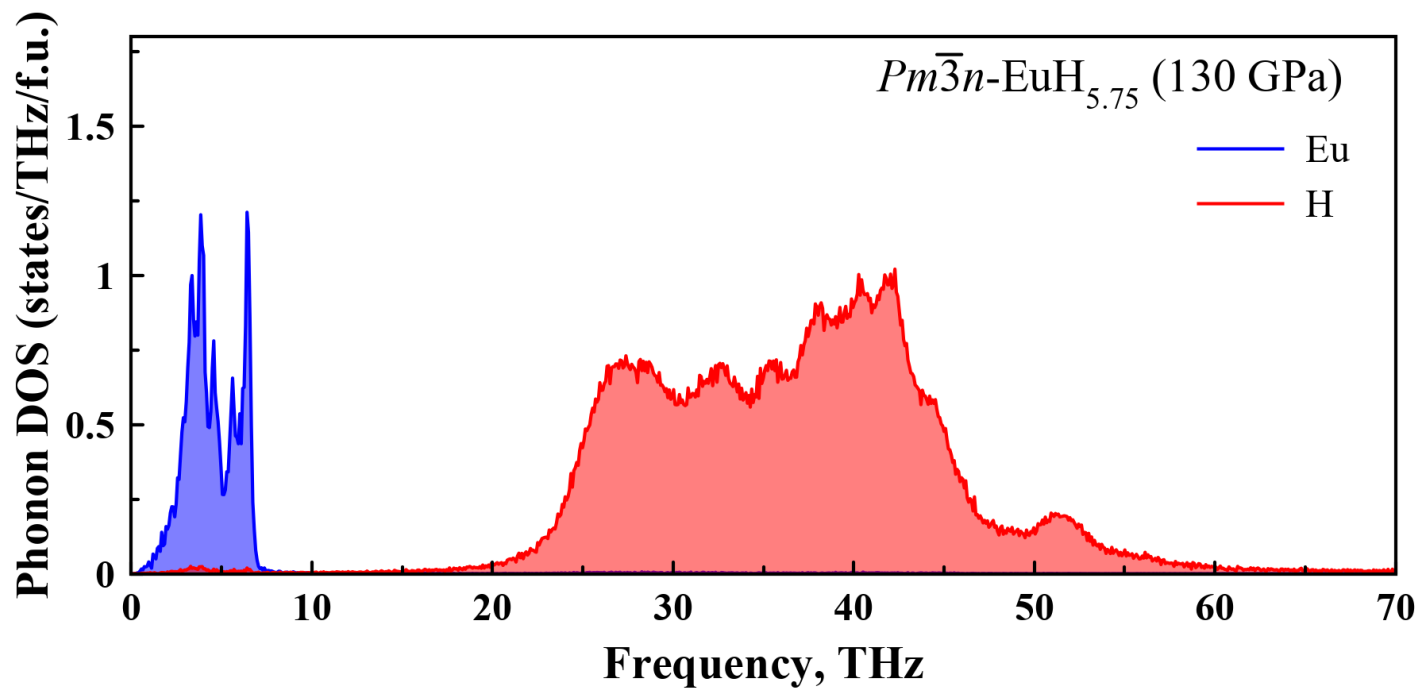

**Figure S31.** Anharmonic phonon density of states of $Pm\bar{3}n$-Eu$_8$H$_{46}$ at 130 GPa and 300 K.



# Elastic Properties of the Eu–H phases

The elastic tensors of the Eu–H phases were calculated using the stress–strain relations:

$$C_{ij} = \frac{\partial \sigma_i}{\partial \eta_j} \quad (S4)$$

where $\sigma_i$ is the $i$th component of the stress tensor, $\eta_j$ is the $j$th component of the strain tensor.

The bulk and shear moduli $B$ and $G$ and Young's modulus $E$ were calculated in GPa via the Voigt–Reuss–Hill averaging.[35,36] Using the obtained values of the elastic moduli, we calculated the velocities of longitudinal and transverse acoustic waves:

$$v_{LA} = \sqrt{\frac{C_{11}}{\rho}}, \quad v_{TA} = \sqrt{\frac{C_{11}-C_{12}}{2\rho}} \quad (S5)$$

where $C_{11}$, $C_{12}$ are the elastic constants, $\rho$ is the density of a compound. The obtained values allow us to estimate the Debye temperature:[36]

$$\vartheta_D = \frac{h}{k_B}\left[\frac{3n}{4\pi}\left(\frac{N_A \cdot \rho}{M}\right)\right]^{\frac{1}{3}} v_m \quad (S6)$$

where $h$, $k_B$, $N_A$ are the Planck, Boltzmann, and Avogadro constants, and $v_m$ is the average velocity of acoustic waves calculated with the following formula:

$$v_m = \left[\frac{1}{3}\left(\frac{2}{v_{TA}^3} + \frac{1}{v_{LA}^3}\right)\right]^{-1/3} \quad (S7)$$

**Table S19.** Elastic and thermodynamic parameters of EuH$_9$ at 130 GPa (SOC+U). To simplify the calculations, ENCUT was reduced to 400–500 eV.

| Parameter | $P6_3/mmc$-EuH$_9$ | $F\bar{4}3m$-EuH$_9$ |
|---|---|---|
| $a$, Å | 3.5450 | 5.009 |
| $c$, Å | 5.9303 | - |
| $V_{DFT}$, Å$^3$ | 32.27 | 31.41 |
| $C_{11}$, GPa | 466 | 561 |
| $C_{12}$, GPa | 198 | 112 |
| $C_{13}$, GPa | 226 | 112 |
| $C_{33}$, GPa | 439 | 561 |
| $C_{44}$, GPa | 143 | 148 |
| $B$, GPa | 297 | 261 |
| $G$, GPa | 131 | 175 |
| $E$, GPa | 343 | 429 |
| Poisson's ratio $\eta$ | 0.307 | 0.226 |
| Debye temperature $\theta_D$, K | 970 | 987 |
| $\omega_{log} = 0.827\theta_D$, K | 802 | 817 |